\documentclass[twocolumn,showpacs]
{revtex4}

\usepackage{latexsym}
\usepackage{bm}

\def\cH{{\cal H}}

\def\cS{{\cal S}}

\def\tr{{\rm tr}}
\def\ket#1{\mid~\!\!\!{#1}~\!\!\rangle}
\def\bra#1{\langle~\!\!{#1}~\!\!\!\mid}
\def\IF{if and only if }
\def\MI{measuring instrument }
\def\mi{measuring instrument}

\def\QM{quantum mechanics }
\def\qm{quantum mechanics}
\def\QMl{quantum-mechanical }

\def\CC{calibration condition }
\def\cc{calibration condition}
\def\PRC{probability reproducibility condition }
\def\prc{probability reproducibility condition}
\def\cR{{\cal R}}
\def\ON{orthonormal }

\def\${\enskip$}
\def\M{measurement }
\def\m{measurement}
\def\Q{quantum }

\def\IN{^\mathbf i}
\def\f{^\mathbf f}

\begin{document}

\title[Kinds of Premeasurement]{A Review of Unitary Quantum Premeasurement Theory\\ An Algebraic Study of Basic Kinds of Premeasurements}

\author{Fedor Herbut}
\affiliation {Serbian Academy of Sciences
and Arts, Knez Mihajlova 35, 11000
Belgrade, Serbia}

\email{fedorh@sanu.ac.rs}

\date{\today}

\begin{abstract}
A detailed theory of \Q pre\M dynamics is presented in which a unitary composite-system operator that contains the relevant object-measuring-instrument interaction brings about the final pre\M state. It does not include collapse, and it does not consider the environment. It is assumed that a discrete degenerate or non-degenerate observable is measured. Pre\M is defined by the calibration condition, which requires that every initially statistically sharp value of the measured observable has to be detected with statistical certainty by the measuring instrument. The entire theory  is derived as a logical consequence of this definition using the standard \Q formalism. The study has a comprehensive coverage, hence the article is actually a topical review. Connection is made with results of other authors, particularly with basic works on pre\m . The article is  a conceptual review, not a historical one.

General exact pre\M is defined in 7 equivalent ways. A step is taken towards complete \M (that includes collapse). Nondemolition pre\m , defined by requiring preservation of any sharp value of the measured observable, is characterized in 10 equivalent ways. Over\m , i. e., a process in which the observable is measured on account of being a function of a finer observable that is actually measured, is discussed. Disentangled pre\m , in which, by definition, to each result corresponds only one pointer-observable state in the final composite-system state, is investigated. Ideal pre\m , a special case of both nondemolition pre\M and disentangled pre\m , is defined, and its most important properties are discussed. Finally, disentangled and entangled pre\m s, in conjunction with nondemolition or demolition pre\m s, are used for classification of all pre\m s. In concluding remarks the omitted aspects of the intricate topic of \Q \M theory are shortly enumerated.
\end{abstract}

\pacs{03.65-w, 03.65 Ca, 03.65 Ta}

\maketitle

\rm
\noindent
{\bf CONTENTS}\\

\noindent I. Introduction: Glance at History; Large Informational Gap; Balance and Dirac Notation; Basic Tools; Basic Aim; Conventions

\noindent II. General       Premeasurement

\noindent III. Towards Complete Measurement

\noindent IV.  Nondemolition Premeasurement

\noindent V. Functions of the Masured Observable;
\noindent
Minimal Premeasurement and Overmeasurement

\noindent VI.  Disentangled Premeasurements

\noindent
VII. Ideal Premeasurements

\noindent
VIII. Subsystem Measurement and Distant Measurement

\noindent IX. Classification

\noindent X. Summing Up the Equivalent Definitions

\noindent XI. Concluding Remarks

\noindent App Proof of an auxiliary algebraic certainty claim\\

\noindent {\bf I. INTRODUCTION}\\

To cast a quick {\it \large glance at history}, one can say that the \QMl theory of pre\M began in the last chapter of von Neumann's celebrated book (von Neumann, 1955). Since then numerous authors gave contributions, but none as many, so it seems to me, as Peka J. Lahti with various coauthors and by himself.  (See his review Lahti, 1993. Some of his articles and those of others will be cited when they overlap with the claims made and proved in the present article.) The efforts of Lahti and coauthors seem to have  culminated in the book by Busch, Lahti, and Mittelstaedt, 1996, which, no doubt, completed the mentioned work of von Neumann in a masterful fashion.

Nevertheless, a {\it\large large informational gap} was left in the wake of this awesome work. The gap is between its highbrow mathematical language and wide generality adopted in the book on the one hand, and the concepts and symbols that are standard in most \QMl textbooks and articles on the other.

Besides, there is the complementarity between the whole and the part well-known even in everyday experience: viewing the wood, you don't see the trees; watching the trees, the wood is out of sight. Comparing Busch, Lahti, and Mittelstaedt 1996 to a magnificent 'wood', the 'trees' would be the generalized observables (so-called positive-operator valued or POV measures; see concluding remark IV in section X), and the ordinary observables could be likened to 'bushes'.

If I am allowed to follow this metaphor a step further, I think that most of us quantum physicists 'live' in the 'bushes'. I think that a {\it \large balanced} approach between the general and the special is desirable, in which one confines oneself to {\bf       pre\M of ordinary observables with purely discrete spectra} and with arbitrarily degenerate or non-degenerate eigenvalues. As to the formalism, it is good to stick to the widely used practical {\it \large Dirac notation}. It follows von Neumann's representation-independent (or abstract) treatment, which is best suited for fundamental \QMl investigations. Besides, unitary evolution operators are made use of. These are the confines of the present topical review, which is written mostly in the way of a research article because almost everything that is claimed is derived from the basic simple definition of       pre\m .

Measurement theory investigates composite systems consisting of the measured object, we will denote it as subsystem A, and the \mi , subsystem B. The {\it\large basic tool} for the treatment of subsystems is the use of  {\bf partial traces}. These have three very practical  accompanying {\bf rules} (much utilized in the present article).

$$\tr_B\Big(Y_AX_{AB}\Big)=Y_A \tr_BX_{AB}\equiv Z_A^{(1)},\eqno{(1a)}$$

$$\tr_B\Big(X_{AB}Y_A\Big)=\Big( \tr_BX_{AB}\Big)Y_A\equiv Z_A^{(2)},\eqno{(1b)}$$
\noindent
where \$Y_A\enskip\Big(=Y_A\otimes I_B\Big)\$ (\$I_B\$ being the identity operator for subsystem B) and \$X_{AB}\$ are arbitrary subsystem and composite-system operators respectively, and the  operators \$Z_A^{(i)}\enskip i=1,2\$, acting (in general nontrivially) in the subsystem state space opposite to that of the partial trace, are defined by the rest in relations (1a,b).
Further, one has
$$\tr_B\Big(Y_BX_{AB}\Big)= \tr_B\Big(X_{AB}Y_B\Big)\equiv Z_A^{(3)}.\eqno{(1c)}$$
We will refer to (1c) as to {\bf 'under-the-partial-trace commutativity'}.

These general {\bf rules} are easily proved
by straightforward evaluation of both sides in an arbitrary pair of complete orthonormal bases  \$\{\ket{k}_A:\forall  k\}\$, \$\{\ket{n}_B:\forall  n\}\$.

The use of partial traces with the rules (1a-c) is wider than \M theory; it can be applied in any \QMl treatment of composite systems.\\

{\it {\Large Literature} To my knowledge, the partial trace was introduced by von Neumann (1955, section 2 of Chapter VI).}\\

The {\it\large basic aim} of this article is to present general       pre\M and its important special case nondemolition pre\M in as many details as possible. The above explained balance has enabled the present author to derive 7 equivalent definitions for general       pre\m , and 10 equivalent definitions for nondemolition pre\m . The rest is a conceptual framework.\\

In pre\M the separate results are not selected out, which would require theoretically some kind of collapse.\\

{\it {\Large Literature} In Busch, Lahti, and Mittelstaedt 1996 the term "pre\m " is used. Complete \M is called "pre\m " with "objectification". The latter term is a synonym of "collapse".}\\

{\bf Complete \m }, which by definition ends in one definite result, consists of two parts: In the first, the measured object interacts in a specific way with the measuring instrument, and specific \Q correlations are created. In the second part, the correlations are 'read' as information about the object on part of the measuring instrument, which thus becomes a 'subject' (it is here a 'technical' term). In the second part, as we know from experience, collapse to one definite result takes place. In the present article only the first part is investigated by applying {\bf unitary evolution operators}, and by treating the \mi s as \QMl systems.

One should keep in mind that unitary \qm , opponents call it sometimes "bare \qm ", is actually the textbook \qm , in which all dynamical changes are expressed via the famous Schr\"odinger equation.

Collapse has been for a long time paradoxical from the point of view of unitary dynamics, because it cannot be derived from the latter in a consistent and expected way (von Neumann, 1955, last section VI.3). It can be done in an unexpected way called relative-state or Everettian theory Everett 1957, Everett 1973 (cf remark II in the Concluding remarks).

By "\m " one usually means complete \M in the literature. This term will be used in the present article when one has both or either of pre\M and complete \M in mind.\\

The present article will use the following {\it\large conventions}. Some physical notions and their mathematical representatives will be used interchangeably throughout, like
'pure state' and 'state vector' (vector of norm one); 'state', 'density operator', and 'state operator'; 'observable' and 'Hermitian operator'; 'compatible' observables and 'commuting' Hermitian operators'; event' and 'projector'. Complete orthonormal bases in a given state space will be called simply 'bases'. Vectors of norm that is not necessarily one will be written overlined; non-overlined vectors will always be of norm one.

Tensor products of vectors will mostly be written without a multiplication sign, but sometimes, for emphasis, such a sign
will be utilized, e. g., \$\ket{\phi}_A\IN\ket{\phi}_B\IN =\ket{\phi}_A\IN
\otimes\ket{\phi}_B\IN \$ (as will be used below). Whenever possible, strings of symbols will begin by numbers that will be followed by the sign "$\times$". We use the convention that if in a string of entities the first is zero, the rest need not be defined; the string is by definition zero.\\

It should be noted that a "necessary and sufficient condition" is another definition. Occasionally also the terms "characterization", "characteristic properties", "criterion", and "condition" will be used as further synonyms of "definition".\\

In order to avoid overburdening the article with mathematics, mathematical presentation in terms of lemmata, theorems etc. will be avoided and boldface "claims", strictly identified via the numbers of the corresponding relations, will be used instead. This seems more appropriate for a physical text. Besides, it is easier for the reader to look up a given relation than to find a lemma or a theorem etc. because the former are all consecutively enumerated, whereas in the latter case the lemmata, theorems etc are usually each separately consecutively enumerated.

Both the proofs and the remarks on literature are written in italics to enable the reader to skip them easily in a first reading.\\

\noindent {\bf II. GENERAL       PREMEASUREMENT}\\

Let subsystem A be the object of \m , and let $$O_A=\sum_ko_kE_A^k,\quad k\not= k'\enskip
\Rightarrow\enskip o_k\not= o_{k'}\eqno{(2a)}$$ be {\bf the unique spectral form} of the measured discrete observable, which may have an infinite purely discrete spectrum and each eigenvalue may have an arbitrary (finite or infinite) degeneracy or lack of it. By 'uniqueness' is meant the non-repetition of the eigenvalues \$\{o_k:\forall  k\}\$ in (2a). Henceforth, we always mean by 'spectral form' the unique one unless otherwise stated. The symbol "$\Rightarrow$" denotes logical implication.

Also the completeness relation \$\sum_kE_A^k=I_A\$ is satisfied.

Let, further, subsystem B be the measuring instrument equipped with a so-called pointer observable
$$P_B=\sum_kp_kF_B^k,\eqno{(2b)}$$ in its spectral form. Also (2b) is accompanied by the corresponding completeness relation \$\sum_kF_B^k=I_B\$.The eigen-projectors \$F_B^k\$ of the pointer observable will be called '{\bf pointer positions}' in the present article . (The term usually applies rather to the eigenvalues \$p_k\$ in the literature, but they will be seen to play an inferior role.)\\

{\it {\Large Literature} We make \$O_A\$ and \$P_B\$ the basic 'players' in the theory following von Neumann; cf von Neumann 1955, end of the third page of section 3. of chapter VI.}\\

The pre\M interaction establishes entanglement between object and measuring instrument in a specific way in the final composite pre\M state. It is investigated in what follows.
The {\bf unitary operator} incorporating the {\bf pre\M interaction} and mapping the initial composite-system state vector \$\ket{\phi}_A\IN\otimes\ket{\phi}_B\IN \$ into the {\bf final state} (at the end of pre\M interaction) we denote by \$U_{AB}\$: $$\forall \ket{\phi}_A\IN\in\cH_A:\quad
\ket{\Phi}_{AB}\f \equiv U_{AB}\Big(\ket{\phi}_A\IN\ket{\phi}_B\IN \Big).\eqno{(3)}$$
Here \$\ket{\phi}_A\IN\$ is an arbitrary initial state vector of the measured system \$A\$, \$\cH_A\$ is the state space (complex separable Hilbert space) of the object, and \$\ket{\phi}_B\IN \$ is {\bf the initial or ready-to-measure state vector} of the instrument. The superscripts refer to the initial and the final state respectively.

The three-tuple \$(\ket{\phi}_B\IN ,P_B,U_{AB})\$ is sometimes called the \$"\MI "\$ in the formalism of unitary \M theory.\\

The degeneracies (or multiplicities) of the  eigenvalues \$o_k\$ of the measured observable may be arbitrary. Also the ranges \$\cR(F_B^k)\$ of the 'pointer positions' \$F_B^k\$ may be degenerate with any degeneracy including the possibility of all being non-degenerate, when one can write \$P_B=\sum_kp_k\ket{k}_B\bra{k}_B\$ (cf (2b)).\\

{\it {\Large Literature} For the last mentioned entirely non-degenerate choice of the pointer positions, Busch, Lahti, and Mittelstaedt (1996)  say that "it is {\bf minimal} in  the sense that it is just sufficient to distinguish the eigenvalues of" the measured observable (2a) (cf subsection III.2.3 in their book, where a different notation is used).}\\

{\bf Premeasurement is defined} by the so-called {\bf \CC}: If the initial state of the object has a sharp (or definite) value of the measured observable, then the final composite-system state has the {\bf corresponding} sharp value of the pointer observable . 'Corresponding' we write as 'having the same index value' \$k\$.

It is firmly established that the \QMl relations have a statistical meaning and are tested on ensembles of equally prepared systems. In particular, the precise {\bf statistical form} of the {\bf \CC} is expressed in terms of the standard {\bf probability formulae}: For any fixed value \$\bar k\$ of the index \$k\$, one has
$$\bra{\phi}_A\IN E_A^{\bar k}\ket{\phi}_A\IN =1\quad\Rightarrow\quad \bra{\Phi}_{AB}\f F_B^{\bar k}\ket{\Phi}_{AB}\f =1,\eqno{(4)}$$  and the final state \$\ket{\Phi}_{AB}\f \$ is given by (3). Equivalent definitions of the calibration condition that are more operational are derived below.

The obvious {\bf physical meaning} of the \CC (4) is that a (statistically) sharp value of the measured observable must be (statistically) sharply detected by the measuring instrument. The \CC is  obviously a necessary requirement for       pre\m . Following Busch, Lahti, and Mittelstaedt 1966 for our restricted choice of observable, we take the \CC as the definition, i. e., as a necessary and sufficient condition, for general exact pre\m .\\

{\it {\Large Literature} The \CC is given in Busch, Lahti, and Mittelstaedt 1996, subsection III.2.3.}\\

The recent ontic breakthrough (Pussey, Barrett, and Rudolph 2012 etc.; cf Herbut 2014a and Leifer 2014) makes it plausible that the state vector is a property of the individual system. Then it is desirable to understand the \CC as a {\bf primarily individual-system} requirement. This means that it may go beyond statistical (and ensemblewise) meaning (relations (4)) towards the individual systems in the following sense. If the \CC were a purely statistical requirement, then it would allow not to be true on some individual systems in any finite ensemble , say on N' of them. But the  relative number N'/N of such exceptions is required to tend to zero as N, the number of systems in the ensemble, tends to infinity. If the \CC  is an individual-system requirement, then it is never allowed to be violated by any individual system in  pre\M .

Nevertheless, the purely statistical notion (4) for pre\M is more suited in view of the fact that all \QMl formulae have, actually, a statistical meaning. One must also be aware that there exist strictly ensemble \m s that cannot be given individual-system meaning. A well-known elementary example is two-slit interference. (When the individual photon hits the second screen, this does not tell much about it. Only an ensemble of photons can reveal interference.) Such \m s are outside the scope of this investigation.\\

To derive an equivalent, more practical, form of (4), we need a useful, general, known, but perhaps not well known, {\bf auxiliary algebraic certainty claim}.\\

An event \$E\$ is certain, i. e., has probability one, in a pure state \$\ket{\psi}\$ \IF the latter is invariant under the action of \$E\$:
$$\bra{\psi}E\ket{\psi}=1 \enskip\Leftrightarrow\enskip E\ket{\psi}=\ket{\psi}.\eqno{(5)}$$ (The symbol "$\Leftrightarrow$" denotes logical implication in both directions.) Proof is given, for the reader's convenience, in the Appendix.\\

Equivalence (5) makes it obvious that the calibration condition can be equivalently defined by the more operational {\bf  invariance form of the \cc }, which is:
$$\ket{\phi}_A\IN =E_A^{\bar k}\ket{\phi}_A\IN
\quad\Rightarrow\quad\ket{\Phi}_{AB}\f =
F_B^{\bar k}\ket{\Phi}_{AB}\f .\eqno{(6)}$$\\

{\it {\Large Literature} The invariance form (6) of the \CC was utilized in Herbut 2014c.}\\

As to the {\bf measuring instrument},
the following {\bf claim } is valid.
The set of all states \$\ket{\phi}_B\IN \$ that satisfy the \CC  in pre\M span a {\bf subspace} \$\cS_B\IN\enskip\Big(\equiv \{all \ket{\phi}_B\IN\}\Big)\$ in the state space \$\cH_B\$ of the measuring instrument in which every state vector is a 'ready-to-measure' state, i. e. $$\ket{\phi}_B\IN\in\cS_B\IN ,\quad \Rightarrow\quad\ket{\phi}_B\IN\in\{all \ket{\phi}_B\IN\}.\eqno{(7)}$$

{\it \small {\Large Proof} \small  is a straightforward consequence of the form (6) of the \CC and the linearity and the continuity of all operations involved. Let \$\ket{\phi}_A\IN =E_A^{\bar k}\ket{\phi}_A\IN\$, and  let \$\ket{\phi}_B\IN \$ and \$\ket{\chi}_B\IN \$ be two initial states of the measuring instrument for which the \CC is valid. Then, for any complex numbers \$\alpha,\beta\$ satisfying
\$|\alpha|^2+|\beta|^2=1\$, (6) implies
$$U_{AB}\Big[\ket{\phi}_A\IN\otimes
\Big(\alpha\ket{\phi}_B\IN
+\beta\ket{\chi}_B\IN \Big)\Big]=$$ $$\alpha F_B^{\bar k}U_{AB}\Big(
\ket{\phi}_A\IN\ket{\phi}_B\IN \Big)+\beta F_B^{\bar k}U_{AB}\Big(
\ket{\phi}_A\IN\ket{\chi}_B\IN \Big)=$$ $$
F_B^{\bar k}U_{AB}\Big[\ket{\phi}_A\IN\otimes
\Big(\alpha\ket{\phi}_B\IN
+\beta\ket{\chi}_B\IN \Big)\Big].$$ Thus, the \CC is valid also for \$(\alpha\ket{\phi}_B\IN
+\beta\ket{\chi}_B\IN ) \$.

Analogously, one proves preservation under the limit process: Let
\$\{(\ket{\phi}_B\IN )^n:n=1,2,\dots ,\infty\}\$ be a sequence of states observing the \CC  that converge to \$\ket{\phi}_B\IN \$, then the \CC is preserved under the limes due to the continuity of both the unitary evolution operator and the projector \$F_B^{\bar k}\$.}\\

The subspace \$\cS_B\IN \$ can be one-dimensional. As an alternative, one can define each pre\M with a fixed initial state \$\ket{\phi}_B\IN \$ disregarding the dimension of \$\cS_B\IN \$. This will be done throughout in this article in order to avoid overburdening the presentation.\\

Now we state and prove the {\bf claim of the dynamical necessary and sufficient condition} for pre\m .

One has  pre\M  {\bf \IF } $$\forall \ket{\phi}_A\IN,\enskip\forall  k:\enskip\Big(F_B^kU_{AB}\Big)\Big(\ket{\phi}_A\IN
\ket{\phi}_B\IN \Big)=$$ $$ \Big(U_{AB}E_A^k\Big)\Big(\ket{\phi}_A\IN
\ket{\phi}_B\IN \Big) \eqno{(8)}$$ is valid.\\

{\it \small One {\Large proves necessity} as follows. For each \$k\$ value the completeness relation \$\sum_{k'}E_A^{k'}=I_A\$,   repeated use of the \CC  in its invariance form (6), and finally orthogonality and idempotency of the \$F_B^k\$ projectors enable one to write $$F_B^kU_{AB}\Big(\ket{\phi}_A\IN\ket{\phi}_B\IN\Big) =$$ $$\sum_{k'}||E_A^{k'}\ket{\phi}_A\IN ||\times F_B^k
U_{AB}\Big[\Big( E_A^{k'}\ket{\phi}_A\IN\Big/ ||E_A^{k'}\ket{\phi}_A\IN ||\Big)
\ket{\phi}_B\IN\Big] =$$ $$\sum_{k'}||E_A^{k'}\ket{\phi}_A\IN ||\times F_B^k
\mathbf{F_B^{k'}}U_{AB}\Big[\Big( E_A^{k'}\ket{\phi}_A\IN\Big/ ||E_A^{k'}\ket{\phi}_A\IN ||\Big)
\ket{\phi}_B\IN\Big] =$$ $$||E_A^k\ket{\phi}_A\IN ||\times F_B^kU_{AB}\Big[\Big( E_A^k\ket{\phi}_A\IN\Big/ ||E_A^k\ket{\phi}_A\IN ||\Big)
\ket{\phi}_B\IN\Big] =$$ $$
||E_A^k\ket{\phi}_A\IN ||\times U_{AB}\Big[\Big( E_A^k\ket{\phi}_A\IN\Big/ ||E_A^k\ket{\phi}_A\IN ||\Big)
\ket{\phi}_B\IN\Big] .$$

After cancellation of \$||E_A^k\ket{\phi}_A\IN ||\$, the claimed relation (8) follows.

Note that the argument covers both the case \$||E_A^k\ket{\phi}_A\IN ||>0\$ and \$||E_A^k\ket{\phi}_A\IN ||=0\$ due to the convention that, when the first factor in a term is zero, then, by definition, the entire term is zero. Namely, in the latter case, it has just been shown that if \$E_A^k\ket{\phi}_A\IN =0\$, then also \$F_B^k\ket{\Phi}_{AB}\f =0\$, i. e., (8) is valid.

To prove {\Large sufficiency}, let (8) be valid, and let \$\ket{\phi}_A\IN =E_A^{\bar k}\ket{\phi}_A\IN\$ be satisfied. Then, one has $$\Big( U_{AB}E_A^{\bar k}\Big) \Big(\ket{\phi}_A\IN\ket{\phi}_B\IN \Big)= \Big(F_B^{\bar k}U_{AB}\Big)
\Big(\ket{\phi}_A\IN\ket{\phi}_B\IN \Big).$$ One can here omit \$E_A^{\bar k}\$ due to the above assumption, and thus the calibration condition in its invariance form (6) \$\ket{\Phi}_{AB}\f =F_B^{\bar k}\ket{\Phi}_{AB}\f \$ is obtained. Hence, we do have       pre\m .}\\

{\it {\Large Literature} The dynamical definition of general       pre\M (8) is introduced and proved in Herbut 2014c.}\\

In an attempt to comprehend the meaning of the dynamical criterion (8) {\bf intuitively}, one may realize that complete \m , which is beyond this study, would collapse the final pre\M state \$\ket{\Phi}_{AB}\f \$ given by (3) into one of its sharp pointer-position projections \$F_B^k \ket{\Phi}_{AB}\f \Big/
||F_B^k\ket{\Phi}_{AB}\f ||\$, and \$E_A^k\ket{\phi}_A\IN\Big/
||E_A^k\ket{\phi}_A\IN ||\$ is the corresponding collapsed form of the initial state. Then (8) says that it amounts to the same whether the collapse takes place at the beginning and then evolution sets in or at the end after the evolution.\\

Important consequences of the dynamical definition (8) of pre\M apply to a connection with complete \m . These consequences are discussed in the next section.\\

To express the {\bf claim of the basis-dynamical characterization} of general       pre\m , we take an arbitrary basis \$\{\ket{k,q_k}_A:\forall q_k\}\$ in each eigen-subspace \$\cR(E_A^k)\$ of the measured observable \$O_A\enskip\Big(=\sum_ko_kE_A^k\Big)\$. Then, the final state (3) is that of a pre\M {\bf  \IF }
$$\forall  k, \forall  q_k:\quad U_{AB}\Big(\ket{k,q_k}_A\ket{\phi}_B\IN \Big)
\in\cR(I_A\otimes F_B^k).\eqno{(9a)}$$\\

{\it \small {\Large Proof of necessity}        is obvious, and so is that of {\Large sufficiency} if one has in mind that an initial state \$\ket{\phi}_A\IN\$ has a sharp value \$o_{\bar k}\$ of \$O_A\$ \IF it can be expanded as \$\ket{\phi}_A\IN =\sum_{q_{\bar k}}\Big(\bra{\bar k,q_{\bar k}}_A
\ket{\phi}_A\IN\Big)\times\ket{\bar k,q_{\bar k}}_A\$ (cf the \CC (6)).}\\

Since the power of the set \$\{\ket{k,q_k}_A\ket{\phi}_B\IN :\forall  q_k\}\$ equals the dimension of \$\cR(E_A^k)\$ and since \$dim\Big(\cR(E_A^k)\Big)\leq dim(\cH_A)
\leq dim(\cH_A)\times dim\Big(\cR(F_B^k)\Big)\$, a unitary operator \$U_{AB}\$ satisfying the basis-dynamical condition can always be constructed. Thus, all discrete observables are, in principle, exactly measurable. (At least there is no algebraic obstacle for this; cf concluding remark VI in section X.)\\

{\it {\Large Literature} In the book Busch, Lahti, and Mittelstaedt 1996 this is proved in Theorem 2.3.1 with the assumption of a pointer observable that is discrete and nondegenerate (\$\forall  k:\enskip dim\Big(\cR(F_B^k)\Big)=1\$).}\\

It is obvious that  the basis-dynamical condition has its equivalent form in the {\bf subspace-dynamical} condition (like the other face of the coin):
The final state \$\ket{\Phi}_{AB}\f\$ is that of an       pre\M \IF each eigen-subspace \$\cR(E_A^k)\$ of the measured observable is taken into the corresponding eigen-subspace \$\cR(I_A\otimes F_B^k)\$ by the unitary evolution operator \$U_{AB}\$. In terms of a system of formulae the characterization has the precise form $$\forall k:\quad U_{AB}\Big[\cR(E_A^k)\otimes\cR(\ket{\phi}_B\IN
\bra{\phi}_B\IN )\Big]\subseteq\cR(I_A\otimes F_B^k),\eqno{(9b)}$$ where by a tensor (or direct) product of subspaces is meant the span of the set of all tensor products of an element from the first and an element from the second factor space.\\

Our next {\bf claim of the \PRC as a necessary and sufficient condition } for pre\M reads: The {\bf probability} of a result \$E_A^k\$ predicted by any initial state \$\ket{\phi}_A\IN\$ of the object {\bf equals the probability} of the corresponding pointer position \$F_B^k\$ in the final composite-system state: $$\forall \ket{\phi}_A\IN,\enskip\forall  k:\enskip \bra{\phi}_A\IN E_A^k\ket{\phi}_A\IN =
\bra{\Phi}_{AB}\f F_B^k\ket{\Phi}_{AB}\f .\eqno{(10)}$$

One should note that one way to put the \CC is to claim that every Kronecker distribution of probability in the sense of LHS(10) equals the corresponding Kronecker probability distribution in the sense of RHS(10). Obviously, the \PRC is valid only if so is the \cc . Though the \PRC apparently requires much more than the \cc , perhaps surprisingly, the former is valid {\bf \IF } so is the latter (both in the statistical sense).\\

{\it {\Large Literature} This is a known fact, cf Busch, Lahti, and Mittelstaedt 1996,  subsection III.1.2).}\\

{\it \small To {\Large prove} that the \CC implies the \prc , we argue as follows. On account of (8), (3), and the idempotency of the projectors, \$RHS(10)\$ equals} $$\bra{\phi}_A\bra{\phi}_B\IN (E_A^k
U_{AB}^{-1})(U_{AB}E_A^k)
\ket{\phi}_A\ket{\phi}_B\IN =LHS(10).$$

{\bf Intuitively}, one may remark that the \PRC (10) displays the kind of information that is transmitted from object to measuring instrument when       \M entanglement in the final state is established.\\

Our next {\bf claim} is another {\bf necessary and sufficient condition} for       pre\m .

{\bf The strong invariance form of the \cc } reads: The  \CC  is valid \IF the pre\M entities \$\{U_{AB},P_B,\ket{\phi}_B\IN \}\$ lead to a final state \$\ket{\Phi}_{AB}\f \$ so that
$$\ket{\phi}_A\IN =E_A^{\bar k}\ket{\phi}_A\IN\quad \Leftrightarrow\quad \ket{\Phi}_{AB}\f = F_B^{\bar k}\ket{\Phi}_{AB}\f \eqno{(11)}$$ holds true.\\

{\small \it Claim (11) {\Large follows} immediately from the \PRC (10) because, according to the latter, $$\bra{\phi}_A\IN E_A^{\bar k}\ket{\phi}_A\IN =1=
\bra{\Phi}_{AB}\f F_B^{\bar k}\ket{\Phi}_{AB}\f $$
(cf (4), (5) and (6)).}\\

The strong invariance form of the \CC implies that the \CC (6) is essentially satisfied also in the time-reversed situation. Hence, the strong invariance form (11) of the \CC  implies a kind of {\bf time reversal symmetry in pre\m }: If we slightly generalize the pre\M concept allowing the initial object+instrument system to be correlated (due to previous interaction that was completely independent of the pre\m ), and we apply time reversal
to the pre\M process, then \$\ket{\Psi}_{AB}\f \$ becomes the initial state, the instrument, subsystem B, is then the measured object, the former pointer observable \$P_B\enskip\Big(=\sum_kp_kF_B^k\Big)\$ is the measured observable, and \$\ket{\phi}_A\IN\ket{\phi}_B\IN \$ the final state. The formerly measured observable \$O_A\enskip\Big(=\sum_ko_kE_A^k\Big)\$ is now the pointer observable with its eigen-projectors \$\{E_A^k:\forall  k\}\$ as the 'pointer positions'.

Now \$\ket{\phi}_A\IN\$ reproduces the relevant information contained in
\$\ket{\Psi}_{AB}\f \$ in terms of the
\PRC $$\forall \ket{\Phi}_{AB}\f ,\enskip\forall  k:\enskip \bra{\phi}_A\IN E_A^k\ket{\phi}_A\IN =
\bra{\Psi}_{AB}\f F_B^k\ket{\Psi}_{AB}\f . \eqno{(12)}$$

{\it {\Large Literature} Peres gave a similar discussion in Peres 1974.}\\

Finally, we turn to the {\bf canonical subsystem-basis expansion criterion} for       pre\m .
The {\bf claim} reads: an object+(measuring instrument) state \$\ket{\Phi}_{AB}\enskip\Big(\equiv U_{AB}
(\ket{\phi}_A\IN\ket{\phi}_B\IN )\Big)\$ is the final state of pre\m , i. e., \$\ket{\Phi}_{AB}=\ket{\Phi}_{AB}\f \$,  {\bf if }, expanded in some eigen-basis \$\{\ket{k,s_k}_B:\forall  k,s_k\}\$ of the pointer observable \$P_B\enskip\Big(=\sum_kp_kF_B^k\Big)\$, in
$$
\ket{\Phi}_{AB}=\sum_k\sum_{s_k}
\overline{\ket{k,s_k}}_A
\ket{k,s_k}_B,\eqno{(13a)}$$ the
'expansion coefficients' (elements of \$\cH_A\$) \$\{\overline{\ket{k,s_k}}_A:\forall k,s_k\}\$ satisfy the following relations:
$$\forall  k:\quad\sum_{s_k}||\overline{\ket{k,s_k}}_A||^2=
\bra{\phi}_A\IN E_A^k\ket{\phi}_A\IN .\eqno{(13b)}$$ One is dealing with the final state of an       pre\M {\bf only if} relations (13a,b) are valid for every eigenbasis of the pointer observable.\\

{\it \small {\Large Proof} of the claim follows immediately from the fact that (13a) and (13b) together are an equivalent formulation of the \prc . Namely,} \$\forall k\$: $$LHS(13b)=||F_B^k\ket{\Phi}_{AB}\f ||^2=\bra{\Phi}_{AB}\f F_B^k
\ket{\Phi}_{AB}\f =RHS(10).$$\\

Note that the 'expansion coefficients' in (13a) are, apart from (13b), arbitrary vectors in \$\cH_A\$. This is what makes the pre\M that we investigate in this section general. Kinds of premeasurement will be specified in the special cases studied further below.\\

{\it {\Large Literature} A somewhat simplified form of the above characterization of pre\M in terms of the form of the final state was obtained by Lahti (1990, relation (13) there).}\\

\noindent {\bf III. TOWARDS COMPLETE MEASUREMENT}\\

To begin with, we specify the explicit expanded form of  the final pre\M state (3) in terms of final complete-\M states \$F_B^k\ket{\Phi}_{AB}\f\Big/||F_B^k\ket{\Phi}_{AB}||\$: $$\ket{\Phi}_{AB}\f =\sum_k(\bra{\phi}_A\IN
E_A^k\ket{\phi}_A\IN )^{1/2}
\times\Big( F_B^k\ket{\Phi}_{AB}\f \Big/
||F_B^k\ket{\Phi}_{AB}\f ||\Big)\eqno{(14)}$$ valid for every \$\ket{\phi}_A\IN\$ (cf (3)). Naturally, it follows from  \$||F_B^k\ket{\Phi}_{AB}\f ||=\Big(
\bra{\Phi}_{AB}\f F_B^k\ket{\Phi}_{AB}\f \Big)^{1/2}\$ and the \PRC (10).

Expansion (14) displays a {\bf connection between pre\M and complete \m }.

The concept of complete \M utilized in (14) is without over\M  (cf section V). In other words, each value of the measured observable is measured {\bf minimally} (more in section V).

{\it {\Large Literature}: The notion of minimal \M was introduced and elaborated in previous work Herbut 1969.}\\

Incidentally, the probability, given by the square of the expansion coefficient in (14), is seen to be {\bf independent} of the kind of       pre\M performed. This is a known (often tacit) textbook claim.\\

To formulate a further important {\bf consequence} of the dynamical condition (8), attention is called to {\bf  concepts relevant for complete \m }.

If \$\bra{\phi}_A\IN E_A^k\ket{\phi}_A\IN >0\$ (non-zero probability), then we refer to the term \$\mathbf{E_A^k\ket{\phi}_A\IN }\$ in the expansion
\$\ket{\phi}_A\IN =\sum_{k'}E_A^{k'}\ket{\phi}_A\IN\$  as the {\bf \$k$-th initial component} (of the object state). Further, we refer to the corresponding final vector \$\mathbf{F_B^k\ket{\Phi}_{AB}\f }\$ in the expnsion \$\ket{\Phi}_{AB}\f=\sum_{k'}F_B^{k'}\ket{\Phi}_{AB}\f\$, due to \$\sum_{k'}F_B^{k'}=I_B\$, as the  {\bf \$k$-th complete-\M final component}. (Note that both are not of norm one in general.)\\

The initial and final components are closely connected. Namely, the dynamical definition (8) implies that {\bf only the \$k$-th initial component  contributes to the \$k$-th complete-\M final component in the  unitary evolution of       pre\m }: $$\forall \ket{\phi}_A\IN,\enskip\forall  k:\quad F_B^k\ket{\Phi}_{AB}\f =
U_{AB}\Big(E_A^k\ket{\phi}_A\IN\otimes
\ket{\phi}_B\IN \Big).\eqno{(15)}$$ Relation (15) is actually (8) rewritten.

Claim (15) is relevant for complete \M of the observable \$O_A\$ in which the result \$o_k\$ is obtained because this process ends in the state \$F_B^k\ket{\Phi}_{AB}\f \Big/
||F_B^k\ket{\Phi}_{AB}\f ||\$ or in a normalized projection of this state by a sub-projector of \$F_B^k\$ in case of over\m .\\

We refer to the initial and the final \$k$-th components {\bf together} as to the \$k$-th {\bf branch} (or channel) of pre\m . This is a concept that applies to the entire pre\M process, not to any fixed moment \$t,\enskip t\IN\leq t\leq t\f\$ in it.

What relation (15) 'tells us' can be put in {\bf intuitive} terms as follows: {\bf The \M process takes place within each branch separately}, independently of the rest of the branches.\\

\noindent {\bf IV. NONDEMOLITION PREMEASUREMENT}\\

{\bf Nondemolition} (synonyms: predictive, repeatable, first kind) {\bf pre\M } is defined as a pre\M satisfying the {\bf additional requirement} that, if the measured observable \$O_A\enskip\Big(=\sum_ko_kE_A^k\Big)\$  has a {\bf sharp value} in the initial state , then, in the final state, the sharp value is {\bf preserved} (it is not demolished).

In the {\bf statistical} language of quantum-mechans the {\bf additional  nondemolition requirement} reads:  $$ \bra{\phi}_A\IN E_A^{\bar k}\ket{\phi}_A\IN =1\quad\Rightarrow\quad \bra{\Phi}_{AB}\f E_A^{\bar k}\ket{\Phi}_{AB}\f =1.\eqno{(16a)}$$ This together with the statistical form of the \CC (4) is the {\bf extended statistical \CC definition} of nondemolition pre\m .

Using the more practical form (5) of certainty, the nondemolition additional condition (16a) can be given the more practical equivalent {\bf invariance form} $$\ket{\phi}_A\IN =E_A^{\bar k}\ket{\phi}_A\IN
\quad\Rightarrow\quad\ket{\Phi}_{AB}\f =
E_A^{\bar k}\ket{\Phi}_{AB}\f .\eqno{(16b)}$$
If joined to the invariance form of the \CC (6) of general       pre\m , then we have the {\bf extended invariance form of the definition} of nondemolition pre\m .\\

We have also {\bf the additional strong invariance requirement}
$$ \ket{\phi}_A\IN =E_A^{\bar k}\ket{\phi}_A\IN \quad\Leftrightarrow\quad
\ket{\Phi}_{AB}\f =E_A^{\bar k}\ket{\Phi}_{AB}\f ,
\eqno{(17)}$$ which together with (11) is the {\bf extended strong invariance definition} of nondemolition pre\m .

The equivalent definitions of nondemolition premeasurement are presented, as far as possible, in an order parallelling that of the definitions of general premeasurement. For this reason, (17) is given here without proof. Proof is supplied below; (17) is a consequence of (20).\\

Since requirement (16b) of nondemolition of the measured eigenvalue of \$O_A\$, joined to the \cc , makes the premeasurement evolution operator \$U_{AB}\$ more specified, it is to be expected that it stands in some additional relation to the eigen-projectors \$E_A^k\$ (with respect to that in the dynamical relation (8)). Indeed, so it is. The following additional {\bf system of dynamical necessary and sufficient conditions} for a general       pre\M to be a {\bf nondemolition} one is valid.

The {\bf claim} of the {\bf additional dynamical condition} goes as follows. A pre\M is a nondemolition one \IF {\bf the evolution operator \$U_{AB}\$ commutes with each eigen-projector \$E_A^k\$} of the measured observable\$O_A\$ when acting on \$\Big(\ket{\phi}_A\IN \ket{\phi}_B\IN \Big)\$. In terms of formulae, a pre\M is a nondemolition pre\M \IF

$$\forall \ket{\phi}_A\IN,\enskip\forall  k:\quad \Big(U_{AB}E_A^k\Big)\Big(\ket{\phi}_A\IN
\ket{\phi}_B\IN \Big)=$$ $$\Big(E_A^kU_{AB}\Big)
\Big(\ket{\phi}_A\IN\ket{\phi}_B\IN \Big) \eqno{(18)}$$ is satisfied.\\

{\it \small To {\Large prove necessity}, let \$\{\ket{k,q_k}_A:\forall  k,\forall  q_k\}\$ be an eigenbasis of the measured observable: $$\forall  k,k',q_k,q_{k'}: \quad E_A^k\ket{k',q_{k'}}_A= \delta_{k,k'}\ket{k,q_k}_A,$$ and let \$\ket{\phi}_B\IN \$ observe the \CC  and the nondemolition requirements (16b) with all pure initial states of the object. Then the eigenvalue relations and the nondemolition requirement (16b) imply $$\forall  k,k',q_{k'}:\qquad \Big(U_{AB}E_A^k\Big)\Big( \ket{k',q_{k'}}_A\ket{\phi}_B\IN \Big)=\delta_{k,k'}\times$$ $$  U_{AB}\Big(\ket{k,q_k}_A\ket{\phi}_B\IN \Big)=
\delta_{k,k'}E_A^k\Big[U_{AB}
\Big(\ket{k,q_k}_A\ket{\phi}_B\IN \Big)\Big].$$

On the other hand, using (16b) again, one has $$\Big(E_A^kU_{AB}\Big) \Big( \ket{k',q_{k'}}_A\ket{\phi}_B\IN \Big)=  E_A^kE_A^{k'}\Big(U_{AB} \ket{k',q_{k'}}_A\ket{\phi}_B\IN \Big)=$$  $$\delta_{k,k'}E_A^kU_{AB}\Big(
\ket{k',q_{k'}}_A\ket{\phi}_B\IN \Big).$$ Since the right-hand sides are equal (if \$k=k'\$, and both are zero otherwise), so are the left-hand sides.

{\Large Sufficiency} is proved in the following way.  Let \$\ket{\phi}_A\IN =E_A^{\bar k}\ket{\phi}_A\IN\$ and let condition (18) be valid. Then
$$U_{AB}\Big(\ket{\phi}_A\IN
\ket{\phi}_B\IN \Big)=E_A^{\bar k}U_{AB}\Big(\ket{\phi}_A\IN
\ket{\phi}_B\IN \Big).$$ Hence, on account of definition (3), also \$\ket{\Phi}_{AB}\f =E_A^{\bar k}\ket{\Phi}_{AB}\f \$, i. e., relation (16b) is satisfied.}\\

The dynamical condition (8) and (18) together  give the {\bf extended dynamical definition} of nondemolition pre\m .\\

Next, the {\bf claim} of the {\bf additional basis-dynamical condition} can be put as follows. A given pre\M is a nondemolition one if and only if,
under the assumptions for claim (9a),  $$\forall  k, \forall  q_k:\quad U_{AB}\Big(\ket{k,q_k}_A\ket{\phi}_B\IN \Big)
\in\cR(E_A^k\otimes I_B)\eqno{(19a)}$$ is satisfied.\\

{\it \small {\Large Proof} is obvious (cf that of (9a)).}\\

Conditions (9a) and (19a) can be  given the form of {\bf one condition}: $$\forall  k, \forall  q_k:\quad U_{AB}\Big(\ket{k,q_k}_A\ket{\phi}_B\IN \Big)
\in\cR(E_A^k\otimes F_B^k).\eqno{(19b)}$$ It is the {\bf extended basis-dynamical definition} of nondemolition pre\m .

Since \$dim\Big(\cR(E_A^k)\Big)
\leq dim\Big(\cR(E_A^k)\otimes\cR(F_B^k)\Big)\$,
the construction of
\$U_{AB}\$ is always possible. Hence, nondemolition pre\M of any discrete observable is, in principle, possible (as far as the algebra of unitary pre\M theory goes, cf concluding remark VI in section X).\\

The 'other face of the coin' of the one relation expressing the basis-dynamical condition (19b), is the {\bf extended subspace-dynamical condition} of nondemolition pre\m . It is: $$\forall  k:\quad U_{AB}\Big[\cR(E_A^k)\otimes\cR(\ket{\phi}_B\IN
\bra{\phi}_B\IN )\Big]\subseteq\cR(E_A^k\otimes F_B^k).\eqno{(19c)}$$\\

There is another {\bf additional system of necessary and sufficient conditions} for a pre\M to be a nondemolition one. It is the {\bf twin-observables condition}.\\

A      pre\M is a nondemolition one \IF its final state satisfies the conditions  $$\forall  \ket{\phi}_A\IN,\enskip\forall  k:\qquad E_A^k\ket{\Phi}_{AB}\f =
F_B^k\ket{\Phi}_{AB}\f \eqno{(20)}$$ (cf (3)), i. e., \IF all pairs of events \$E_A^k\$ and \$F_B^k\$ are  so-called {\bf twin projectors} in it, and hence the measured observable \$O_A\$ and the pointer observable \$P_B\$ are {\bf twin observables} in it.\\

{\it {\Large Literature} Twin observables were introduced in Herbut and Vuji\v{c}i\'{c} 1976, and elaborated in Herbut 2002. (They are presented in detail in the unpublished review Herbut 2014b.)}\\

{\it \small {\Large Proof of Necessity} follows from \$\sum_kE_A^k=I_A\$, (3), the \CC (6), and the nondemolition condition (16b): $$\ket{\Phi}_{AB}\f =$$ $$
\sum_k||E_A^k\ket{\phi}_A\IN ||\times U_{AB}\Big[\Big(E_A^k\ket{\phi}_A\IN \Big/
||E_A^k\ket{\phi}_A\IN ||\Big)\otimes
\ket{\phi}_B\IN \Big]=$$  $$
\sum_k||E_A^k\ket{\phi}_A\IN ||\times  E_A^kF_B^kU_{AB}\Big[\Big(E_A^k\ket{\phi}_A\IN
\Big/||E_A^k\ket{\phi}_A\IN ||\Big)
\otimes\ket{\phi}_B\IN \Big].$$ The rest is obvious.

To prove {\Large sufficiency}, let \$\ket{\phi}_A\IN =E_A^{\bar k}\ket{\phi}_A\IN\$. On accont of (6), one has \$\ket{\Phi}_{AB}\f =F_B^{\bar k}\ket{\Phi}_{AB}\f \$. Further, (20) implies \$\ket{\Phi}_{AB}\f =E_A^{\bar k}\ket{\Phi}_{AB}\f \$, i. e., the sharp value of the measured observable is not demolished ((16b) is valid).}\\

One should note that there is no condition for general pre\M that would be parallel to the 'twin-observables condition' (20).

Condition (20) has important {\bf consequences}.
First of all, it implies the additional strong invariance form of the \CC (17) in an obvious way (if one has in mind the strong invariance form of the \CC (11)). But there is more.

{\bf A)} In the final subsystem states (reduced density operators) $$\rho_A\f \equiv
\tr_B\Big(\ket{\Phi}_{AB}\f \bra{\Phi}_{AB}\f \Big)\quad \mbox{and}\quad\rho_B\f \equiv
\tr_A\Big(\ket{\Phi}_{AB}\f \bra{\Phi}_{AB}\f \Big)$$ there is {\bf no coherence} with respect to the events \$E_A^k\$ and \$F_B^k\$ respectively: $$\forall \ket{\phi}_A\IN\in\cH_A:\quad\rho_A\f =
\sum_kE_A^k\rho_A\f E_A^k,\eqno{(21a)}$$
$$\forall \ket{\phi}_A\IN         :\quad\rho_B\f =
\sum_kF_B^k\rho_B\f F_B^k.\eqno{(21b)}$$

{\bf B)} The following commutations are valid: \$\forall \ket{\phi}_A\IN         ,\enskip\forall  k:$ $$ [E_A^k,\rho_A\f ]=0,\qquad [F_B^k,\rho_B\f ]=0.\eqno{(22a,b)}$$\\

{\it \small  {\Large Proof} of claim {\Large A)}: Since \$\sum_kF_B^k=I_B\$, one has $$\rho_A\f =
\sum_{k,k'}\tr_B\Big(F_B^k\ket{\Phi}_{AB}\f
\bra{\Phi}_{AB}\f F_B^{k'}\Big).$$ Utilizing under-the-partial-trace commutativity (1c) twice, and orthogonality of the projectors, one further obtains
$$\rho_A\f =
\sum_{k,k'}\tr_B\Big[\Big({\bf F_B^{k'}}F_B^k\Big)
\ket{\Phi}_{AB}\f \bra{\Phi}_{AB}\f \Big]=$$ $$\sum_k\tr_B\Big(F_B^k\ket{\Phi}_{AB}\f
\bra{\Phi}_{AB}\f F_B^k\Big).$$ The twin relations (20) and the fact that opposite-subsystem operators can be taken outside the partial trace (cf (1a,b)) enable one to write further $$\rho_A\f =
\sum_k\tr_B\Big(E_A^k\ket{\Phi}_{AB}\f
\bra{\Phi}_{AB}\f E_A^k\Big)=$$ $$
\sum_kE_A^k\Big[\tr_B\Big(\ket{\Phi}_{AB}\f
\bra{\Phi}_{AB}\f \Big)\Big]E_A^k=
\sum_kE_A^k\rho_A\f E_A^k.$$

One proves symmetrically, exchanging the roles of the two subsystems and of \$F_B^k\$ and \$E_A^k\$, that $$\rho_B\f =\sum_kF_B^k\rho_B\f F_B^k.$$ {\Large Proof} of claim {\Large B)}. It is straightforward to see
 that claims A and B are equivalent.}.\\

One should note that the relation \$\rho_A\f =\sum_{k,k'}E_A^k\rho_A\f E_A^{k'}\$
is an identity on account of the completeness relation \$\sum_kE_A^k=I_A\$. Due to the fact that none of the {\bf coherence terms} \$E_A^k\rho_A\f E_A^{k'}\enskip k\not= k'\$ is non-zero (cf (22a)), one says that there is {\bf no coherence} in the final subsystem state \$\rho_A\f \$ as far as the measured observable \$O_A\$ is concerned. So it is symmetrically in \$\rho_B\$. This means that one is dealing with {\bf improper mixtures} (D'Espagnat, 1976) \$\rho_A\f =\sum_k[\tr(\rho_A\f E_A^k)]\times
\Big(E_A^k\rho_A\f E_A^k\Big/[\tr(\rho_A\f E_A^k)]\Big)\$,
etc.\\

{\it {\Large Literature} The twin relations (20) are characterized in Busch and Lahti 1996
, sections 4 and 5, as "strong correlations", which are an extremal case of "correlations". The latter are applicable also to general       pre\m s. They are described in detail in subsections III.3.1-III.3.4 .}\\

Another characterization of nondemolition pre\M that has no parrallel claim for general       pre\m s is the
{\bf claim of repeatability}. It asserts that nondemolition complete \M can be equivalently defined by requiring that, for every initial state of the measured object,  immediate repetition of the same complete \M necessarily (with statistical necessity) gives the same result: $$\forall \ket{\phi}_A\IN ,\enskip\forall  k,\enskip
\bra{\phi}_A\IN E_A^k\ket{\phi}_A\IN >0:$$  $$
\Big(\bra{\Phi}_{AB}\f F_B^k\Big)E_A^k
\Big(F_B^k\ket{\Phi}_{AB}\f \Big)
\Big/||F_B^k\ket{\Phi}_{AB}\f ||^2=1.\eqno{(23a)}$$

Making use of (5), (23a) can be equivalently written as $$\forall \ket{\phi}_A\IN ,\enskip\forall  k,\enskip
\bra{\phi}_A\IN E_A^k\ket{\phi}_A\IN >0:$$  $$
E_A^kF_B^k\ket{\Phi}_{AB}\f =F_B^k\ket{\Phi}_{AB}\f .
\eqno{(23b)}$$\\

{\it \small Proof of {\Large necessity} If the pre\M is a nondemolition one, then the twin observables condition (20) is valid. Hence $$\forall \ket{\phi}_A\IN ,\enskip\forall  k,\enskip \bra{\phi}_A\IN\enskip E_A^k\ket{\phi}_A\IN >0:$$ $$
E_A^kF_B^k\ket{\Phi}_{AB}\f =E_A^kE_A^k
\ket{\Phi}_{AB}\f =
F_B^k\ket{\Phi}_{AB}\f ,$$ i. e. (23b) is satisfied.

Proof of {\Large sufficiency} Let \$\ket{\phi}_A\IN =E_A^{\bar k}\ket{\phi}_A\IN\$ be valid. Then,
utilizing the \CC (6) twice, and making use of (23b), we have \$\ket{\Phi}_{AB}\f =E_A^{\bar k}F_B^{\bar k}\ket{\Phi}_{AB}\f =E_A^{\bar k}\ket{\Phi}_{AB}\f \$.
Thus, the additional nondemolition condition (16b) for nondemolition pre\M is satisfied.}\\

{\it {\Large Literature} The 'repeatability condition' is one of the ways how Pauli defined nondemolition \M in Pauli 1933.}\\

As it was stated at the beginning of this section, also the term "predictive pre\m " is used as a synonym of "nondemolition pre\m " in unitary \M theory of discrete observables. It suggests that the complete \M result predicts the result of an immediately repeated identical \m . "Retrodictive" or "retrospective" \M is then a synonym for "demolition \m ".\\

There is yet another {\bf additional necessary and sufficient condition} for nondemolition pre\m . We shall call it the {\bf extended \prc }. It reads
$$\forall \ket{\phi}_A\IN ,\enskip\forall k:\quad\bra{\phi}_A\IN E_A^k\ket{\phi}_A\IN =\bra{\Phi}_{AB}\f E_A^k\ket{\Phi}_{AB}\f .\eqno{(24)}$$

{\it \small Proof of  {\Large necessity} follows from the \PRC (10): $$\forall  \ket{\phi}_A\IN ,\enskip\forall  k:\quad\bra{\phi}_A\IN E_A^k\ket{\phi}_A\IN =\bra{\Phi}_{AB}\f F_B^k\ket{\Phi}_{AB}\f .$$ Taking into account the validity of the twin-observables condition (20), this immediately becomes (24).

To prove {\Large sufficiency}, we take \$\ket{\phi}_A\IN =E_A^{\bar k}\ket{\phi}_A\IN \$. Then (24) implies \$\bra{\Phi}_{AB}\f E_A^{\bar k}\ket{\Phi}_{AB}\f =1\$, or, equivalently on account of (5), \$E_A^{\bar k}\ket{\Phi}_{AB}\f =
\ket{\Phi}_{AB}\f\$. Thus, the sharp value is preserved, and we have nondemolition pre\m .\\

{\it {\Large Literature} Pauli (1933) gave the condition at issue the following catchy {\bf physical meaning}: "The initial state and the final state of premeasurement predict the same probabilities for the \M in question.

In the book by Busch, Lahti, and Mittelstaedt (1996), which treats a wider class of \M processes, the 'repeatability condition' and that of the 'extended \prc ' are not equivalent; the former is a special case of the latter (see subsections III.3.5 and III.3.6 there). Both in Pauli 1933 and in Busch, Lahti, and Mittelstaedt 1996 the latter \M is called "of the first kind". And so it is in most of the literature. (The term is abandoned in this review because "first kind" does not suggest the essence of the condition.)}\\

\rm
Finally we give two canonical-expansion criteria for nondemolition pre\m . Let us first specify the {\bf subsystem-basis canonical expansion one}.\\

{\bf A)} A {\bf sufficient} additional condition for nondemolition pe\M reads: A final state \$\ket{\Phi}_{AB}\f \$ of pre\M (cf (3)) is that of a nondemolition one {\bf if} there exists an {\bf  eigen-basis} \$\{\ket{k,s_k}_B:\forall  k,s_k\}\$ of the pointer observable \$P_B\$ (cf (2b)) such that, when \$\ket{\Phi}_{AB}\f \$ is expanded in it, each nonzero 'expansion coefficient' (vector in \$\cH_A\$) is an eigen-vector of the measured observable \$O_A\enskip\Big(=\sum_ko_kE_A^k\Big)\$ with the same \$k$-value:  $$\forall\ket{\phi}_A\IN         : \quad \ket{\Phi}_{AB}\f =\sum_{k,s_k}
\overline{\ket{k,s_k}}_A\ket{k,s_k}_B\eqno{(25a)}$$
implies
$$\forall  k,s_k:\quad
E_A^k\overline{\ket{k,s_k}}_A=
\overline{\ket{k,s_k}}_A.
\eqno{(25b)}$$

{\bf B)} The final state \$\ket{\Phi}_{AB}\f \$ is that of nondemolition pre\M {\bf only if} (25a,b) is valid for its expansion in {\bf every} eigen-basis of the pointer observable.\\

{\it \small To prove {\Large necessity }, we point out that (25a) is equivalent to the partial scalar product $$\forall  k:\quad\overline{\ket{k,s_k}}_A=    \bra{k,s_k}_B\ket{\Phi}_{AB}\f.$$ Applying \$E_A^k\$ to both sides, and utilizing the twin-observables condition (20), we obtain $$\forall  k:\quad E_A^k\overline{\ket{k,s_k}}_A=
\Big(\bra{k,s_k}_BF_B^k\Big)\ket{\Phi}_{AB}\f =
\overline{\ket{k,s_k}}_A.$$ (Since "partial scalar product'' is not used often in the literature, the reader may be unfamiliar with it. Perhaps he (or she) should read Appendix A in Herbut 2014b.)

To prove {\Large sufficiency}, we assume the validity of (25a,b), and we apply \$E_A^{\bar k}\$ and alternatively \$F_B^{\bar k}\$ to (25a) for an arbitrary fixed value \$k=\bar k\$. Then the twin-observables definition (20) follows.}\\

Another criterion for a pre\M to be a nondemolition one is, what we shall call, the {\bf twin-correlated canonical Schmidt decomposition} criterion. It is in terms of a special kind of a canonical Schmidt or a bi-orthonormal decomposition with positive numerical expansion coefficients of the final state \$\ket{\Phi}_{AB}\f \$.\\

{\bf A)} A pre\M is a nondemolition one {\bf if } there exists a canonical Schmidt decomposition
$$\forall \ket{\phi}_A\IN         :\quad
\ket{\Phi}_{AB}\f =\sum_kr_k^{1/2}\sum_{s_k}
\ket{k,s_k}_A\ket{k,s_k}_B,\eqno{(26)}$$ where \$\{\ket{k,s_k}_B:\forall  k,\forall  s_k\}\$ are {\bf simultaneous} eigen-vectors of the pointer observable \$P_B\enskip\Big(=\sum_kp_kF_B^k\Big)\$, i. e., $$\forall  k:\quad F_B^k=\sum_{s_k}\ket{k,s_k}_B\bra{k,s_k}_B,$$
and the final measuring-instrument state operator (reduced density operator) \$\rho_B\f \enskip\Big[\equiv\tr_A\Big(
\ket{\Phi}_{AB}\f \bra{\Phi}_{AB}\f \Big)\Big]\$
spanning the range \$\cR(\rho_B\f )\$.\\

Note that symmetrically \$\{\ket{k,s_k}_A:\forall  k,\forall  s_k\}\$ are simultaneous eigen-vectors of the measured observable \$O_A\enskip\Big(=\sum_ko_kE_A^k\Big)\$ and the final object state operator (reduced density operator) \$\rho_A\f\enskip\Big[\equiv\tr_B\Big(\ket{\Phi}_{AB}\f
\bra{\Phi}_{AB}\f \Big)\Big]\$ spanning the range \$\cR(\rho_A\f )\$.

Actually, (26) is a subsystem-basis expansion with respect to the instrument-subsystem sub-basis \$\{\ket{k,s_k}_B:\forall  k,\forall  s_k\}\$ (spanning \$\cR(\rho_B\f )\$). Hence, the above common eigen-sub-basis \$\{\ket{k,s_k}_A:\forall  k,\forall  s_k\}\$  for \$O_A\$ and \$\rho_A\f\$ of the object subsystem is uniquely determined by the state \$\ket{\Phi}_{AB}\f \$. (The former is determined via the antiunitary so-called "correlation operator" \$U_a\$ inherent in the composite state. Decomposition (26) is called  twin-correlated canonical Schmidt decomposition due to the twin-observables condition (20).  More about this special kind of Schmidt canonical expansion see in section 5 of Herbut 2014b.)\\

{\bf B)} A pre\M is a nondemolition one {\bf only if} the twin-correlated canonical Schmidt decomposition (26) is valid for {\bf every} common eigen-bases of \$P_B\$ and \$\rho_B\f \$ that span the range of \$\rho_B\f \$.\\

{\it \small Proof of {\Large sufficiency} is immediately obtained by noticing that (26) is a special case of (25a,b). Proof of {\Large  necessity} also follows from this remark and the properties of any twin-correlated canonical Schmidt decomposition. (If in doubt, consult Herbut 2014b.)}\\

{\it {\Large Literature} Twin-correlated canonical Schmidt decompositions were introduced into \QM by von Neumann (1955, see section 2 in chapter VI, in particular pp. 434-436). Von Neumann did not utilize our term.

The antiunitary correlation operator was introduced in Herbut and Vuji\v{c}i\'{c} 1976. The canonical Schmidt (or biorthogonal) decomposition of a bipartite state vector (with and without the explicit correlation operator \$U_a\$, but without twin correlation) was reviewed in subsection 2.1 of Herbut 2007a.

The twin-correlated canonical Schmidt decomposition criterion for a kind of pre\M was investigated also by Beltrametti, Cassinelli, and Lahti (1990). They referred to it as to "strong correlations premeasurement".}\\

Some founders of \QM and many foundationally oriented physicists consider only nondemolition pre\M for {\bf individual} \Q systems. The reason is, of course, the fact that unless one can check the result obtained (repeatability), the individual-system physical meaning of the result of pre\M is doubtful.

Nevertheless, nowadays we are witnessing an ever-increasing permeation of physics by information theory (see e. g. Vedral, 2006). General       pre\m , treated in section II, which includes besides nondemolition also demolition pre\m , transmits information from system to measuring apparatus. In pre\M this is apparent for individual systems if the initial state has a sharp value of the measured observable; otherwise only probability is transmitted (the \prc ), and it is an ensemble phenomenon. But pre\M underlies, in some way, complete \m , and the latter transmits individual-system information. Besides, pre\M theory primarily concerns ensembles. For these reasons thorough investigation of pre\M more general than the nondemolition one in section II is made the basis of this review.\\

\noindent {\bf V. FUNCTIONS OF THE MEASURED OBSERVABLE; MINIMAL PREMEASUREMENT AND OVERMEASUREMENT}\\

We denote by \$f(\dots )\$ any single-valued real function, i. e., any map of the real axis into itself. It determines a Hermitian {\bf operator function} \$\bar O\$ of any given Hermitian operator \$O_A\enskip\Big(=\sum_ko_kE_A^k\Big)\$ of the object subsystem in spectral form as follows: $$\bar O_A\equiv f(O_A)\equiv  \sum_kf(o_k)E_A^k=\sum_lo_lE_A
^l, \eqno{(27a)}$$ where the second spectral form is, by definition,  unique, i. e., it is the same as the first spectral decomposition , but rewritten in the unique form. It satisfies \$l\not=
l'\enskip\Rightarrow o_l\not= o_{l'}\$. In general, the first spectral form is not unique.

The inverse function \$f^{-1}(\dots )\$ is, in general, multi-valued, i. e., its images are sets: \$\forall  l:\enskip f^{-1}(o_l)=\{o_k:f(o_k)=o_l\}\$. We can omit the eigenvalues and keep only their indices because they are in a one-to-one relation (in the unique spectral form). Hence, one can write \$\forall  l:\enskip l=f(k)\$, \$l\$ being the index of \$o_l=f(o_k)\$, and \$\forall  l:\enskip f^{-1}(l)=\{k:f(k)=l\}\$.\\

Parallelly, one defines a {\bf corresponding pointer observable} $$\bar P_B\equiv \sum_lp_lF_B^l,\eqno{(27b)}$$ where $$\forall  l:\quad F_B^l\equiv \sum_{k\in f^{-1}(l)}F_B^k,\eqno{(27c)}$$ and \$\{p_l:\forall l\}\$ are arbitrary distinct real numbers.\\

Whenever the function \$f(\dots)\$ in (27a-c) is {\bf not} one-to-one, i. e., whenever  this function is singular, the \M of \$\bar O_A\$ that is obtained due to a \M of \$O_A\$, is called {\bf over\m } of \$\bar O_A\$. If \$O_A\$ is a complete observable, i. e., if all its eigenvalues are non-degenerate, then \$\bar O_A\$ is said to be {\bf overmeasured maximally} (or just "measured maximally" cf Herbut, 1969).\\

{\it {\Large Literature} In von Neumann 1955 it is mostly assumed that all observables are overmeasured maximally (cf p. 348 not far from the beginning of section 1 of Chapter V).}\\

If an observable is measured so that it is not overmeasured, then one says that it is measured {\bf minimally} (a term introduced in previous work Herbut, 1969).\\

{\it {\Large Literature} It was, actually, L\"uders (1951) who introduced minimal pre\M  via ideal pre\m , though he did not call it so. The way I see it, it was an important development in unitary \M theory after von Neumann. (Unfortunately, some physicists still labor under the illusion that,if there is any \M other than maximal, then it is only ideal \M as L\"uders introduced it.)}\\

The opposite concept of over\M is {\bf under\m }. If, in the notation used above, the observable \$\bar O_A\$ is measured minimally, then \$O_A\$ (cf (27a)) is undermeasured. The best known example of under\M is that of values in a continuous spectrum. The latter is, as a rule, an interval. One breaks it up into a set-theoretical sum of smaller (non-overlapping) intervals. One evaluates the spectral measures of these subintervals and, utilizing them, one  defines a discrete observable. Its \M undermeasures the continuous spectrum. As it is well known, the values of a continuous spectrum cannot be exactly measured. (There are no eigen-projectors corresponding to them.)\\

{\it {\Large Literature} Von Neumann has claimed that a continuous spectrum is normally measured undermeasuring it by a discrete observable as just described (cf von Neumann 1955, chapter III, section 3. p. 220). His term for under\M is "\M with only limited accuracy". Regarding the continuous spectrum, see also Ozawa (1984).}\\

Now we state and prove {\bf the basic  claims on over\m }:

For {\bf every function} \$f(\dots )\$ the following claims are valid.

{\bf A)} If \$\ket{\phi}_B\IN \$ is calibration-condition-satisfying for the pre\M of \$O_A\$, then so it is also for that of \$\bar O_B\equiv f(O_A)\$ in terms of the pointer observable \$\bar P_B\$ (cf (27a-c)). Remembering the dynamical definition of general \m , this claim can read:
$$\forall\ket{\phi}_A\IN ,\enskip\forall k:\enskip
F_B^kU_{AB}\Big(\ket{\phi}_A\IN\ket{\phi}_B\IN\Big) = U_{AB}E_A^k\Big(\ket{\phi}_A\IN\ket{\phi}_B\IN\Big)
\enskip
\Rightarrow$$ $$\forall l\enskip\Big(\equiv f(k)\Big):\enskip F_B^lU_{AB} \Big(\ket{\phi}_A\IN\ket{\phi}_B\IN\Big)= U_{AB}E_A^l  \Big(\ket{\phi}_A\IN\ket{\phi}_B\IN\Big).\eqno{(28a)}$$

{\bf B)} If the pre\M of \$O_A\$ is a {\bf nondemolition} one, then {\bf so is} that of its function \$\bar O_A\equiv f(O_A)\$. Having in mind the definition of nondemolition pre\M in terms of twin observables (20), this additional claim can be put as follows:
$$\forall\ket{\phi}_A\IN ,\enskip\forall k:\quad F_B^k\ket{\Phi}_{AB}\f =E_A^k\ket{\Phi}_{AB}\f\quad
\Rightarrow$$  $$\forall l\enskip\Big(\equiv f(k)\Big),\quad F_B^l\ket{\Phi}_{AB}\f =E_A^l
\ket{\Phi}_{AB}\f .\eqno{(28b)}$$

{\it \small To {\Large prove} {\Large claim A}, we argue as follows. On account of  the assumption that the \CC (6) is valid for the pre\M of \$O_A\$, one has $$\forall \ket{\phi}_A\IN :\quad\ket{\Phi}_{AB}\f \equiv U_{AB}(\ket{\phi}_A\IN\ket{\phi}_B\IN )=$$ $$
\sum_k||E_A^k\ket{\phi}_A\IN ||\times \mathbf{F_B^k } U_{AB}\Big[\Big(E_A^k\ket{\phi}_A\IN\Big/
||E_A^k\ket{\phi}_A\IN ||\Big)\otimes\ket{\phi}_B\IN \Big].$$ Hence, $$\forall \ket{\phi}_A\IN :\quad\ket{\Phi}_{AB}\f =\sum_kF_B^k U_{AB}\Big(E_A^k\ket{\phi}_A\IN\ket{\phi}_B\IN \Big).\eqno{(aux)}$$

Further, we rewrite this as two sums, and we take \$\ket{\phi}_A\IN =E_A^{\bar l}\ket{\phi}_A\IN\$. Then, we replace \$\ket{\phi}_A\IN\$ by \$\mathbf{E_A^{\bar l}}\ket{\phi}_A\IN\$ in the second sum, obtaining $$\ket{\Phi}_{AB}\f =\sum_{k\in f^{-1}(\bar l)}F_B^k U_{AB}\Big(E_A^k
\ket{\phi}_A\IN\ket{\phi}_B\IN \Big)
\mathbf{+}$$
$$\sum_{k\notin f^{-1}(\bar l)}F_B^k U_{AB}\Big(E_A^k\mathbf{E_A^{\bar l}}\ket{\phi}_A\IN
\ket{\phi}_B\IN \Big).$$
Next, we take into account the facts that \$E_A^{\bar l}E_A^k=E_A^k\$ if \$k\in f^{-1}(\bar l)\$, \$E_A^kE_A^{\bar l}=0\$ if \$k\notin f^{-1}(\bar l)\$ and correspondingly (cf (27a-c)) \$F_B^{\bar l}F_B^k=F_B^k\$ if \$k\in f^{-1}(\bar l)\$, \$F_B^{\bar l}F_B^k=0\$ if \$k\notin f^{-1}(\bar l)\$. We thus obtain
$$\ket{\Phi}_{AB}\f =\mathbf{F_B^{\bar l}}\sum_{
k\in f^{-1}(\bar l)}F_B^kU_{AB} \Big(E_A^k\ket{\phi}_A\IN\ket{\phi}_B\IN \Big)=$$ $$F_B^{\bar l}\sum_\mathbf{k}F_B^kU_{AB} \Big(E_A^k\ket{\phi}_A\IN\ket{\phi}_B\IN \Big).$$
(We were able to include the terms for \$k\notin f^{-1}(\bar l)\$ because they are zero.)
Finally, on account of the above general auxiliary relation (aux), we have in our special case $$\ket{\Phi}_{AB}\f =F_B^{\bar l}\ket{\Phi}_{AB}\f .$$ Therefore, the \CC for \$\bar O_A\$ is valid (cf (6)), and we are dealing with a        pre\M of this observable.\\

{\Large Proof of claim B} We assume that we have nondemolition pre\M of \$O_A\$, and we utilize the twin-relations definition (20) for this.. Then, on account of (27a-c),  $$\forall  l:\quad F_B^l
\ket{\Phi}_{AB}\f =\Big(\sum_{k\in f^{-1}(l)}
F_B^k\Big)\ket{\Phi}_{AB}\f =$$ $$\Big(\sum_{k\in f^{-1}(l)}\mathbf{E_A^k}\Big)\ket{\Phi}_{AB}\f =E_A^l
\ket{\Phi}_{AB}\f$$ follows for an arbitrary initial state of the object. We see that the pre\M is nondemolition also for that of \$\bar O_A\$ because the twin relations (20) are valid also for this observable.}\\

\noindent {\bf VI. DISENTANGLED PREMEASUREMENTS}\\

Now the question of inverting the described functional relation \$\bar O_A\equiv f(O_A)\$ (cf (27a-c)) arises. For any answer there is a need for a more precise terminology. We have seen that the eigenvalues do not play an essential role in pre\m ; only the eigen-projectors and their indices do.\\

Having in mind the relation between the indices \$k\$ and \$l\$ (cf (27a-c)), one says, by {\bf definition}, that \$\bar O_A\$ is {\bf coarser} than \$O_A\$, or that the former is a {\bf coarsening} of the latter. The same relation can be expressed also by the terms: \$O_A\$ is {\bf finer} than
\$\bar O_A\$, or the former is a {\bf refinement} of the latter.\\

If a coarsening \$\bar O_A\equiv f(O_A)\$ (cf (27a-c)) is mathematically given, and one measures the coarser observable \$\bar O_A\$ minimally, one may wonder if this means anything for the finer observable \$O_A\$. Clearly, the answer is: one has an {\bf under\M } of  \$O_A\$. This would belong to approximate \M theory, which is outside the scope of this article.\\

It may happen that a given pre\M of an observable \$\bar O_A\$ is actually a pre\M of one of its refinements, i. e., that \$\bar O_A=f(O_A)\$ is valid, and that \$\bar O_A\$ is overmeasured. If this is the case, then, as it was explained, there must exist also a corresponding refinement of the pointer observable \$P_B\$ with co-indexed eigen-projectors, and a subspace of the subspace spanned by all the calibration-condition-satisfying states \$\ket{\phi}_B\IN \$, the unit vectors in which are calibration-condition-satisfying for the refinement. Evidently, it is not always easy to find out if a given observable \$\bar O_A\enskip\Big(=f(O_A)\Big)\$ is actually overmeasured.\\

We now turn to a class of cases in which, though \$O_A\$ itself may have refinements, its given \M is certainly not an overmeasurement, i. e., it is a {\bf minimal \m }.

By {\bf definition}, one has a {\bf disentangled or uncorrelated  pre\M } of an observable \$O_A\enskip\Big(=\sum_ko_kE_A^k\Big)\$ if there exists an eigen-sub-basis \$\{\ket{\phi}_B^k:\forall  k,\ket{\phi}_B^k=F_B^k\ket{\phi}_B^k
\}\$ of the pointer observable \$P_B\$ (cf (2b)), and for each \$k\$ value
{\bf only one eigenvector} \$\ket{\phi}_B^k\$ from this sub-basis appears in the canonical subsystem-basis expansion form (13a) of the final pre\M state. Equivalently, for every initial state \$\ket{\phi}_A\IN\$ of the object, the final state \$\ket{\Phi}_{AB}\f =U_{AB}\Big(\ket{\phi}_A\IN \ket{\phi}_B\IN \Big)\$ can be expanded in this sub-basis:
$$\forall \ket{\phi}_A\IN         :\quad
\ket{\Phi}_{AB}\f =\sum_k\overline{
\ket{\phi}_A^{\mathbf{f},k}}\ket{\phi}_B^k,\eqno{(29a)}$$ where \$\overline{\ket{\phi}_A^{\mathbf{f},k}}\$ are vectors in \$\cH_A\$. Then, equivalently $$\forall  k:\quad\overline{
\ket{\phi}_A^{\mathbf{f},k}}=\bra{\phi}_B^k \ket{\Phi}_{AB}\f ,\eqno{(29b)}$$ the right-hand sides being partial scalar products (cf possibly Appendix A in Herbut 2014b).\\

The term 'disentangled' (introduced by Schr\"odinger, 1935) applies to the {\bf uncorrelated} (disentangled) vectors \$\overline{\ket{\phi}_A^{\mathbf{f},k}}\ket{\phi}_B^k\$ in the final state (29a).\\

Disentangled pre\M can be of the nondemolition kind, when the result of \M is preserved throughout the branches \$\forall  k:\enskip\overline{
\ket{\phi}_A^{\mathbf{f},k}}=E_A^k\overline{
\ket{\phi}_A^{\mathbf{f},k}}\$, or of the demolition kind if the preservation fails for some  \$k\$ value. Among the latter, there is an interesting possibility: the set of 'expansion coefficients' \$\{\forall k:\overline{\ket{\phi}_A^{\mathbf{f},k}}\}\$ in (28a) may turn out {\bf orthogonal} \$\forall  \ket{\phi}_A\IN\$ . Then they determine another observable of subsystem A with sharp values in the selective final states \$F_B^k\ket{\Phi}_{AB}\f \Big/ ||F_B^k\ket{\Phi}_{AB}\f ||\$ possibly independently of the initial state.

It is worth mentioning that, whereas most \M properties are valid (or not valid) for each branch separately; the property at issue is a pre\M property, valid for the entirety of the final state.\\

{\it {\Large Literature} This case is referred to as "strong state correlation" in Busch, Lahti, and Mittelstaedt 1996 (subsection III.3.2).}\\

A  {\bf sufficient condition for disentangled pre\M } appears in nondemolition pre\M of a {\bf complete} observable \$O_A\$: \$O_A=\sum_ko_k\ket{k}_A\bra{k}_A,\enskip k\not= k'\enskip\Rightarrow\enskip o_k\not= o_{k'}\$. Then
the preservation of \$E_A^k=\ket{k}_A\bra{k}_A\$ requirement implies that in each complete-\M branch  \$F_B^k\ket{\Phi}_{AB}\f\$ only \$\ket{k}_A\$ can appear (cf (20)): \$\ket{\Phi}_{AB}\f =
\sum_k\ket{k}_A\otimes
\overline{\ket{\phi}_B^{\mathbf{f},k}}\$. (The state vector \$\ket{k}_A\$, as any state vector, cannot be entangled.)\\

In many discussions of the paradox of the emergence of complete \M in \qm , one confines oneself to the simplified case of disentangled \m . (Most likely, one keeps unitary dynamics as simple as possible in order not to cloud the issue of the paradox.)\\

Another {\bf sufficient condition for disentangled pre\M } is {\bf non-degeneracy of all 'pointer positions'} for obvious reasons (cf in section 2 the passage next to the one in which relation (3) is).\\

{\it {\Large Literature} In Busch, Lahti, and Mittelstaedt 1996 this case is referred to as "the pointer observable is minimal in the sense that it is just sufficient to distinguish between the eigenvalues" of the measured observable (in passage next to (7) in III.2.3).}\\

Let us next state and prove {\bf the basic property} of disentangled \m s. It goes as follows.

If a given pre\M of an observable \$O_A\$ is {\bf disentangled}, then the evolution operator \$U_{AB}\$ {\bf can be replaced} by a set of partial isomorphisms {\bf \$\{U_A^k:\forall  k\}\$} each mapping the range \$\cR(E_A^k)\$ of the corresponding eigen-projector \$E_A^k\$ (cf (1a)) into \$\cH_A\$ isometrically, i. e., linearly and preserving scalar products.
The replacement is performed {\bf so that}: $$\forall  \ket{\phi}_A\IN         :\quad\ket{\Phi}_{AB}^\mathbf f\equiv  U_{AB}
(\ket{\phi}_A\IN\ket{\phi}_B\IN )=$$  $$
\sum_k
(U_A^kE_A^k\ket{\phi}_A\IN )
\otimes\ket{\phi}_B^k\eqno{(30)}$$ is valid.\\

One should note that the operators \$\{U_A^k:\forall  k\}\$, being partial isometries {\bf can be extended into a unitary operator} in \$\cH_A\$, and
they map \ON vectors into \ON ones.\\

{\it \small To {\Large prove} the claim, let us take a complete \ON eigen-basis \$\{\ket{k,q_k}_A:\forall  k,q_k\}\$ of \$O_A\$, and expand \$\ket{\phi}_A\IN =\sum_{k,q_k}\bra{k,q_k}_A
\ket{\phi}_A\IN\times\ket{k,q_k}_A\$. Next, we evaluate the final pre\M state using this expansion.
$$U_{AB}(\ket{\phi}_A\IN\ket{\phi}_B\IN )=
\sum_{k,q_k}\bra{k,q_k}_A\ket{\phi}_A\IN\times
U_{AB}(\ket{k,q_k}_A\ket{\phi}_B\IN ).$$  According to the \CC (6), each final state \$U_{AB}(\ket{k,q_k}_A\ket{\phi}_B\IN )\$ must be invariant under the action of  \$F_B^k\$ on the one hand, and, according to the definition of disentangled pre\m , only one of the given eigen-vectors \$\ket{\phi}_B^k\$, independent of the \$q_k\$ values, can appear in the expansion. Hence, \$U_{AB}\$, being unitary, maps, for each value of \$ k\$, the \ON sub-basis \$\{\ket{k,q_k}_A\ket{\phi}_B\IN :\forall q_k\}\$
into some other \ON sub-basis
\$\{\ket{k,q_k}_A'\ket{\phi}_B^k:\forall q_k\}\$ in \$\cH_A\otimes\cH_B\$. (Note that the state vectors  \$\ket{\phi}_B^k\$, as any state vectors,  cannot be entangled.)
It follows from the orthonormality of \$\{\ket{k,q_k}_A'\ket{\phi}_B^k:\forall q_k\}\$ that also the set \$\{\ket{k,q_{k}}_A':\forall q_k\}\$ is orthonormal.

Each pair of \ON sub-bases determines a partial isometry \$\forall k,\enskip U_A^k\$: \$\forall  q_k:\enskip U_A^k\ket{k,q_k}_A=\ket{k,q_k}_A'\$. Hence, we can write $$U_{AB}(\ket{\phi}_A\IN\ket{\phi}_B\IN )=
\sum_{k,q_k}\bra{k,q_k}_A\ket{\phi}_A\IN\times
(U_A^k\ket{k,q_k}_A)\ket{\phi}_B^k=$$ $$
\sum_k[(U_A^kE_A^k\ket{\phi}_A\IN )
\otimes\ket{\phi}_B^k].$$}\\

The characterization of disentangled pre\M by (30) is the {\bf essence} of this kind of pre\m . The operators \$\{U_A^kE_A^k:\forall  k\}\$ are called {\bf state transformers} because, in disentangled complete \M with the result \$o_k\$, they transform the initial state \$\ket{\phi}_A\IN\$ into \$U_A^kE_A^k\ket{\phi}_A\IN\Big/||U_A^kE_A^k
\ket{\phi}_A\IN ||\$ (cf Herbut, 2004).\\

{\it {\Large Literature} The state transformers in relation (30) are also called  Kraus operators because they have been introduced (as far as I know) by Kraus (1983). The above claim is stated and proved as Theorem 3.2 in Lahti 1990. In Busch, Lahti, and Mittelstaedt 1996 claim (30) is presented in subsection III.2.3 .}\\

As it is easily seen, any disentangled pre\M is a {\bf nondemolition} one \IF $$\forall  k:\quad
U_A^kE_A^k=E_A^kU_A^kE_A^k;\eqno{(31)}$$
otherwise, it is a demolition pre\M (cf (16b)).\\

\noindent {\bf VII. IDEAL PREMEASUREMENTS}\\

The simplest among disentangled pre\m s are the {\bf ideal pre\m s}. They can be defined in several equivalent ways. The ones that are most used are the following.\\

{\bf (I)} \$\forall  k:\enskip U_A^k=I_k\$, where \$I_k\$ is the identity operator in \$\cH_A\$ restricted to the range \$\cR(E_A^k)\$ of \$E_A^k\$ (cf (30)): $$\forall  \ket{\phi}_A\IN         :\quad \ket{\Phi}_{AB}^\mathbf f =
\sum_k(E_A^k\ket{\phi}_A\IN )
\otimes\ket{\phi}_B^k.\eqno{(32)}$$

Expansion (32) is obviously a twin-correlated canonical Schmidt (or bi-orthogonal) decomposition (having the measured observable and the pointer observable as twin observables in mind, cf (20) and (26)). We may call it shortly the {\bf canonical-final-state definition} of ideal pre\m .\\

{\bf (II)} The definition that ensues  may be called the {\bf L\"uders change-of -state one}. It says that that the general final state \$\rho_A\f\$ of the object in ideal pre\M of an observable \$O_A\enskip\Big(=\sum_ko_kE_A^k\Big)\$
is given by: $$\forall\ket{\phi}_A\IN :\enskip \rho_A\f\equiv\tr_B\Big(\ket{\Phi}_{AB}\f \bra{\Phi}_{AB}\f\Big)
=\sum_kE_A^k\ket{\phi}_A\IN \bra{\phi}_A\IN E_A^k.\eqno{(33a)}$$

One can rewrite (33a) as $$\forall\ket{\phi}_A\IN :\enskip \rho_A\f =$$  $$\sum_k \bra{\phi}_A\IN E_A^k\ket{\phi}_A\IN\Big( E_A^k\ket{\phi}_A\IN \bra{\phi}_A\IN E_A^k\Big)\Big/\tr\Big( E_A^k\ket{\phi}_A\IN \bra{\phi}_A\IN E_A^k\Big),\eqno{(33b)}$$ as seen if one applies, in the denominator, under-the-trace commutativity, idempotency, and if one evaluates the trace in a basis containing \$\ket{\phi}_A\IN\$.

In this way it is seen that the {\bf final complete-\M states} are the pure states $$\forall k,\enskip\bra{\phi}_A\IN E_A^k\ket{\phi}_A\IN >0:\quad E_A^k\ket{\phi}_A\IN\Big/||E_A^k\ket{\phi}_A\IN ||,\eqno{(33c)}$$ where \$\bra{\phi}_A\IN E_A^k\ket{\phi}_A\IN\$ are, of course, the probabilities in the improper mixture (33b) (cf D'Espagnat 1976).\\

{\bf (III)} The third definition may be called the {\bf strongly extended \cc }. It reads: Every initial state that has a sharp value of the measured observable does not change at all in ideal pre\m : $$\ket{\phi}_A\IN =E_A^{\bar k}\ket{\phi}_A\IN\quad\Rightarrow\quad
\rho_A\f=\ket{\phi}_A\IN \bra{\phi}_A\IN .\eqno{(34a)}$$\\

This definition of ideal pre\M is in line with our first definitions of general       pre\M (6) and nondemolition pre\M (16b). Comparing (34a) with (16b), it is obvious that {\bf ideal pre\M is a special case of nondemolition pre\m }.

One should note that the first definition of ideal pre\M is not additional to that of general       pre\m ; it defines the pre\M in question by itself completely. The second and third definitions, on the contrary, are additional requirements because pre\M cannot be defined only by changes in the object subsystem.\\

{\it \small We give an {\Large 'in-circle'  proof} (to be distinguished from a circular one) of the equivalence of the three definitions.

The {\Large first implies the second}: Substituting the final pre\M state from (32) both in its ket and bra forms, one obtains
$$\rho_A\f\equiv\tr_B\Big(\ket{\Phi}_{AB}\f
\bra{\Phi}_{AB}\f\Big)=$$  $$\sum_{k,k'}
E_A^k\ket{\phi}_A\IN\bra{\phi}_A\IN E_A^{k'}
\Big(\tr_B(\ket{\phi}_B^k\bra{\phi}_B^{k'} )\Big)=$$
$$\sum_k
E_A^k\ket{\phi}_A\IN\bra{\phi}_A\IN
E_A^k=RHS(33a).$$

The {\Large second implies the third}:
Substituting \$\ket{\phi}_A\IN=E_A^{\bar k}\ket{\phi}_A\IN\$ in (33a) immediately lads to \$\rho_A\f =\ket{\phi}_A\IN\bra{\phi}_A\IN\$.\\

The {\Large third implies the first}.
The third definition (34a) can be completed into $$\ket{\phi}_A\IN =E_A^{\bar k}\ket{\phi}_A\IN \quad\Rightarrow\quad
\ket{\Phi}_{AB}\f =\ket{\phi}_A\IN\ket{\phi}_B^{\bar k}\eqno{(34b)}$$
with $$\ket{\phi}_B^{\bar k}=F_B^{\bar k}\ket{\phi}_B^{\bar k}.\eqno{(34c)}$$

Relations (34b,c) imply that each complete-\M branch
\$E_A^k\ket{\phi}_A\IN\Big/||E_A^k\ket{\phi}_A\IN ||\$ in the decomposition \$\ket{\phi}_A\IN =\sum_k
E_A^k\ket{\phi}_A\IN\$ gives a disentangled component in the final state. Hence, the final pre\M state is disentangled. Having in mind \$I_A=\sum_kE_A^k\$, and (34b) with (34c), one obtains:
$$\ket{\Phi}_{AB}\f =\sum_kU_{AB}\Big((E_A^k\ket{\phi}_A\IN )\ket{\phi_B}\IN\Big) =$$  $$
\sum_k||E_A^k\ket{\phi}_A\IN ||\times
U_{AB}\Big[\Big(E_A^k\ket{\phi}_A\IN\Big/
||E_A^k\ket{\phi}_A\IN ||\Big)
\ket{\phi_B}\IN\Big]=$$ $$
\sum_k||E_A^k\ket{\phi}_A\IN ||\times
\Big(E_A^k\ket{\phi}_A\IN\Big/
||E_A^k\ket{\phi}_A\IN ||\Big)\mathbf{
\ket{\phi_B}^k}=$$  $$
\sum_k(E_A^k\ket{\phi}_A\IN )\ket{\phi_B}^k$$ (cf (32)).}\\

{\it {\Large Literature} The change of state (33a-c) was postulated by L\"uders (1951) (cf Messiah, 1961, or Cohen-Tannoudji, Diu, and Laloe, 1977). For this reason ideal \M is sometimes called L\"uders \m . If the measured observable is complete (no degenerate eigenvlues), then the L\"uders change of state coincides with that  postulated by von Neumann (1955) as his process 1. Therefore ideal \M is sometimes called also von Neumann-L\"uders \m .

Among others, also the present author studied some derivations of the L\"uders change of state Herbut (1969), Herbut (1974), Herbut (2007b). Also Khrennikov (2009a, 2008, and 2009b) has paid great attention to ideal \m .

Incidentally, the postulate of L\"uders (33a-c) was called 'minimal \m ' in Herbut, 1969 to stress the fact that it has introduced a difference with respect to von Neumann (1955), who restricted his discussion to  maximal over\M of degenerate observables, i. e., \M of complete observables whose function the initial observable is.}\\

Ideal \M is not very  important in practice because it cannot be achieved in direct \m . The characteristic property (34a) contradicts the empirical fact that in every direct interaction at least one quantum of action has to be exchanged, i. e., some change in the state has to be brought about. But this \M is of great theoretical importance. This will be elaborated elsewhere.\\

\noindent {\bf VIII. SUBSYSTEM MEASUREMENT AND DISTANT MEASUREMENT}\\

In this section the main results of a recent article (Herbut 2014c) are shortly reviewed to complete the picture on basic kinds of \m .

 Let us envisage a  composite \Q system consisting of subsystems \$A_1\$ and \$A_2\$ and being initially in an, in general, entangled state \$\ket{\phi}_{A_1A_2}\IN\$. Further, let us performs an arbitrary exact subsystem pre\m , i. e.,,pre\M of a subsystem observable, e. g., \$O_{A_2}\enskip\Big(=\sum_ko_kE_{A_2}^k\Big)\$,
on subsystem \$A_2\$. We call subsystem \$A_2\$ the nearby subsystem, and subsystem \$A_1\$ the distant one. (The spatial connotation is here metaphorical; the terms are actually dynamical.)

Let the pre\M end up in the final pre\M state $$\ket{\Phi}_{A_1A_2B}\f \equiv U_{A_1}U_{A_2B}\Big(\ket{\phi}_{A_1A_2}\IN
\ket{\phi}_B\IN\Big),\eqno{(35a)}$$ \$U_{A_1}\$ being the unitary evolution operator of the dynamically isolated distant subsystem. Then the subsystem pre\M exerts {\bf no influence} on the distant  subsystem \$A_1\$ that is 'untouched' by the \M interaction.

Actually, even more is true. It is a known fact that any unitary change to the nearby subsystem \$A_2\$, with or without an ancilla \$A_3\$, does not have any influence on the state of a dynamically isolated subsystem  \$A_1\$. (For details, see section 4 in Herbut 2014c.)

If, on the other hand, one considers the final state $$F_B^k\ket{\Phi}_{A_1A_2B}\f
\Big/||F_B^k\ket{\Phi}_{A_1A_2B}\f ||,\eqno{(35b)}$$
of a complete \m , then there is, in general, a {\bf change} in the state of the 'interactionally untouched' distant subsystem \$A_1\$, which is {\bf due to the entanglement} in the initial state \$\ket{\phi }_{A_1A_2}\IN\$.

Then, the {\bf claim} is that the final distant-subsystem state in question has the form: $$\rho_{A_1}^{\mathbf{f},k}= U_{A_1}\Big(\rho_{A_1}(E_{A_2}^k)\Big)U_{A_1}^{\dag},
\eqno{(36a)}$$ where by $$\rho_{A_1}(E_{A_2}^k)\equiv$$  $$\tr_{A_2}\Big(
\ket{\phi}_{A_1A_2}^\mathbf{i}
\bra{\phi}_{A_1A_2}^\mathbf{i}E_{A_2}^k
\Big)\Big/ \tr\Big(\ket{\phi}_{A_1A_2}^\mathbf{i}
\bra{\phi}_{A_1A_2}^\mathbf{i}E_{A_2}^k\Big)\eqno{(36b)}$$ is denoted the {\bf conditional state} of the distant subsystem \$A_1\$ under the condition of the ideal occurrence of the event \$E_{A_2}^k\$ in the composite-system state \$\ket{\phi}_{A_1A_2}^\mathbf{i}\$ instantaneously at the initial moment. This interpretation of (36b) is obvious if one rewrites the relation (utilizing idempotency and under-the-partial-trace commutativity (1c)) in the equivalent form $$\rho_{A_1}(E_{A_2}^k)=$$  $$  \tr_{A_2}\Big(E_{A_2}^k
\ket{\phi}_{A_1A_2}^\mathbf{i}
\bra{\phi}_{A_1A_2}^\mathbf{i}E_{A_2}^k
\Big)\Big/ ||E_{A_2}^k\ket{\phi}_{A_1A_2}^\mathbf{i}||^2, \eqno{(36c)}$$ utilizing the trivial equality $$
\tr\Big(\ket{\phi}_{A_1A_2}^\mathbf{i}
\bra{\phi}_{A_1A_2}^\mathbf{i}E_{A_2}^k\Big)=
||E_{A_2}^k\ket{\phi}_{A_1A_2}^\mathbf{i}||^2\eqno{(37)}$$
and having in mind relation (33c) for ideal \m .
(The lengthy proof of claim (36a,b) is given in the Appendix of Herbut 2014c.)

On account of the fact that the unitary interaction operator \$U_{A_2B}\$ has dropped out of the final  expression (36a,b), it is an important corollary of the result that, whatever the kind of complete \M performed on \$A_2\$, the change caused to \$A_1\$ is the {\bf same as if the subsystem \M were ideal}.\\

The {\bf change of state due to ideal complete \M } is easily evaluated on account of relation (32) and the \PRC \$||F_B^k\ket{\Phi}_{A_1A_2B}\f ||=
||E_{A_2}^k\ket{\phi}_{A_1A2}\IN ||\$ (cf (10)):
$$F_B^k\ket{\Phi}_{A_1A_2B}\f\Big/
||F_B^k\ket{\Phi}_{A_1A_2B}\f ||=$$ $$U_{A_1}
E_{A_2}^k\ket{\phi}_{A_1A2}\IN\ket{\phi}_B^k\Big/
||E_{A_2}^k\ket{\phi}_{A_1A2}\IN ||.\eqno{(38)}$$

As to the state of the distant subsystem \$A_1\$, one has:
$$\rho_{A_1}^{\mathbf{f},k}\equiv\tr_{A_2B}\Big[
\Big(F_B^k\ket{\Phi}_{A_1A_2B}\f\Big/
||F_B^k\ket{\Phi}_{A_1A_2B}\f ||\Big)$$  $$
\Big(\bra{\Phi}_{A_1A_2B}\f F_B^k\Big/
||F_B^k\ket{\Phi}_{A_1A_2B}\f ||\Big)\Big].$$ Further, substituting here (38) and taking into account (37), one finally obtains
 $$\rho_{A_1}^{\mathbf{f},k}=U_{A_1}\Big[\tr_{A_2}\Big(E_{A_2}^k
 \ket{\phi}_{A_1A2}\IN
\bra{\phi}_{A_1A2}\IN E_{A_2}^k\Big)\Big/$$ $$||E_{A_2}^k\ket{\phi}_{A_1A2}\IN ||^2
\Big]U_{A_1}^{-1}.\eqno{(39)}$$\\

Now we come to {\bf distant \m }. We assume that initially we have a twin-observables relation $$\forall k:\quad E_{A_1}^k\ket{\phi}_{A_1A_2}=
E_{A_2}^k\ket{\phi}_{A_1A_2},\eqno{(40)}$$ where \$E_{A_1}^k\$ are the eigen-projectors of a distant-subsystem observable \$O_{A_1}\enskip\Big(=
\sum_k\bar o_kE_{A_1}^k\Big)\$. (The twin-observables criterion for nondemolition observables (20) is an example for (40)).)

Making use of (40), relation (39) implies $$\rho_{A_1}^{\mathbf{f},k}=
U_{A_1}\Big[\tr_{A_2}\Big(E_{A_1}^k
 \ket{\phi}_{A_1A2}\IN
\bra{\phi}_{A_1A2}\IN E_{A_1}^k\Big)\Big/$$ $$ ||E_{A_1}^k\ket{\phi}_{A_1A_2}\IN ||^2\Big]
U_{A_1}^{-1}.\eqno{(41)}$$

In view of (33c) for ideal \m , relation (41) gives the final distant-subsystem state  due to the instantaneous ideal occurrence of the (twin) distant subsystem event \$E_{A_1}^k\$ in the state \$\ket{\phi}_{A_1A_2}\IN\$ at the initial moment.

Comparing (39) and (41), one can see that both the instantaneous ideal occurrence of the event \$E_{A_2}^k\$ on the nearby subsystem and that of the event \$E_{A_1}^k\$ on the distant subsystem lead to the same final state \$\rho_{A_1}^{\mathbf{f},k}\$ of the distant subsystem \$A_1\$. In other words, one can say that the instantaneous ideal occurrence of \$E_{A_2}^k\$ in the state \$\ket{\phi}_{A_1A_2}\IN\$ at the initial moment gives rise to the instantaneous ideal occurrence of the distant twin event \$E_{A_1}^k\$ in the same state at the same moment.

This was introduced and called {\bf distant \M } a long time ago (Herbut and Vuji\v ci\' c 1976). But now we have the recent (above mentioned) result that any exact nearby-subsystem complete \M leads to the same final state of the distant subsystem as nearby-subsystem ideal complete \m . Hence, one can say that, if twin observables are involved (cf (40)), any exact subsystem \M gives rise to a change of state in the opposite subsystem which is the same as caused by instantaneous ideal measurement of the corresponding distant twin event at the initial moment in the given composite-system state \$\ket{\phi}_{A_1A_2}\IN\$ with entanglement. We can keep the term "distant \m" for this more general \m .

One should note that as long as the nearby-subsystem complete \M is ideal, it gives the same change of the {\bf global} (composite) state as the analogous ideal complete \M on the distant subsystem. If the former complete \M is more general than ideal, then its effect coincides with that of ideal complete \M on the distant subsystem {\bf only locally} on the distant subsystem.\\

Distant \M plays a natural role in the paradoxical so-called Einstein-Podolsky-Rosen (EPR) phenomenon.
If a bipartite state
vector \$\ket{\phi}_{A_1A_2}\IN\$ allows distant
\M of two mutually incompoatible
observables (non-commuting operators)
\$O_{A_2}\$ and \$\bar O_{A_2}\$, \$[O_{A_2},\bar O_{A_2}]\not= 0\$, then we say
that we are dealing with an {\bf EPR
state} (following the seminal
Einstein-Podolsky-Rosen 1935 article).\\

A very simple example of an EPR state is the well known {\bf singlet two-particle spin state}
$$\ket{\phi}_{A_1A_2}\IN\equiv (1/2)^{1/2}\Big(
(\ket{+}_{A_1}\ket{-}_{A_2}-\ket{-}_{A_1}
\ket{+}_{A_2})\Big),
\eqno{(42)}$$ where \$+\$ and \$-\$
denote spin-up and spin-down respectively
{\bf along any fixed axis}. One can see, in obvious notation, that the oppositely oriented spin-projection operators are twin observables:
$$s_{\vec k_0,A_1}\ket{\phi}_{A_1A_2}\IN =
s_{\vec (-k_0),A_2}\ket{\phi}_{A_1A_2}\IN ,\eqno{(43)}$$ where the unit vector \$\vec k_0\$ is arbitrary. One can suitably
choose \$\vec k_0\$ either along the positive  z-axis or along the positive x-axis.

If one performs an ideal complete \M of the spin projection along the z-axis in a subsystem \M on the nearby subsystem \$A_2\$, it is easily seen that the  composite final state is, e. g., \$\ket{z,+}_{A_1}\ket{z,-}_{A_2}\$, and the (this time pure) state of the distant subsystem \$A_1\$ is \$\ket{z,+}_{A_1}\$. Analogously, one can obtain by distant complete \m , e. g., \$\ket{x,+}_{A_1}\$.

Einstein et al. pointed out that in transition from (42) to the final state of distant complete \M, the mentioned result of opposite spin projection along the same axis was brought about in a distant action without interaction ( a "spooky" action), which could not be reconciled with basic physical ideas that reigned outside \qm . (More in section 6.2 of Herbit 2014b or in Bohm 1952, chapter 22, section 15. and further.)

One can find articles in the literature
in which all entangled bipartite states
are called EPR states. Perhaps because any entanglement allows distant complete \M (with intuitively "spooky" action).\\

\noindent {\bf VIII. CLASSIFICATION}\\

In the {\bf classification} that follows we disregard over\m s because in them the measured observable is a function of a finer observable, and the \M is just a consequence of the minimal \M of the latter.\\

{\bf Five kinds of final complete-\M components of minimal \m s} can be distinguished, and they are displayed in the ARRAY below, and denoted by {\bf M}$_{\mathbf{x,y}}\$. They are (in reading order on the Diagram):

1) the {\bf ideal ones, M}$_{\mathbf{1,1};\mathbf{a}},\ $ characterized by {\bf preserving} the sharp-value component {\bf states} (cf (34a));

2) the {\bf nondemolition non-ideal disentangled ones, M}$_{\mathbf{1,1};\mathbf{b}},\$ which do not preserve the sharp-value component states, but they {\bf preserve the sharp values} themselves (cf (16b)), and they {\bf map} the initial pure states \$\ket{\phi}_A\IN\$ {\bf into pure states} $$U_A^k\Big(E_A^k \ket{\phi}_A\IN\Big/||E_A^k \ket{\phi}_A\IN||\Big),\quad U_A^k\not= I_A;\eqno{(44)}$$\\

3) the {\bf nondemolition entangled ones, M}$_{\mathbf{1,2}}\$, which also {\bf preserve} the mentioned {\bf sharp values}, but they {\bf map} the initial pure states {\bf into mixtures} (improper or second-kind ones cf D'Espagnat, 1976)
$$\rho_A\f \equiv\tr_B\Big[\Big(U_{AB}(\ket{\phi}_A\IN
\ket{\phi}_B\IN )\Big)
\Big((\bra{\phi}_A\IN\bra{\phi}_B\IN )U_{AB}^{\dag}\Big)\Big].
\eqno{(45)}$$ In other words, {\bf redundant entanglement} is created between the subsystems \$A\$ and \$B\$ ('redundant' regarding the pre\m ).

4) the {\bf demolition disentangled ones, M}$_{\mathbf{2,1}},\$ which {\bf do not preserve the sharp values}, but they do {\bf map} the intial states \$\ket{\phi}_A\IN\$ {\bf into pure states} given by (44); and finally

5) the {\bf demolition entangled ones, M}$_{\mathbf{2,2}},\$ which {\bf neither preserve the sharp values}, nor do they map the pure states into pure states. They map the former into improper mixtures given by (45), i. e., they create {\bf redundant entanglement}.\\

$\qquad\qquad$\textbf{ARRAY} (with rows and columns)\\

$\qquad\qquad\qquad\quad$\textsc{disentangled}:
$\qquad$\textsc{entangled}:\\

\noindent\textsc{nondemolition}:$\qquad$ M$_{\mathbf{1,1;a,b}}\qquad\qquad\qquad$M$_{\mathbf{1,2}}\qquad\qquad$\\

$\quad$\textsc{demolition}:$\qquad$ M$_{\mathbf{2,1}} \qquad\qquad\qquad\quad$M$_{\mathbf{2,2}}$\\

To understand the {\bf ARRAY}, one must take into account that a {\bf kind of \M M}$_{\mathbf{x,y}}$ is defined by the row \$\mathbf{x}\$ and the column \$\mathbf{y}\$. In the special case of {\bf M}$_{\mathbf{1,1}}$, one has two kinds: {\bf M}$_{\mathbf{1,1};\mathbf{a}}$, which is {\bf ideal \m }, and {\bf M}$_{\mathbf{1,1};\mathbf{b}}$, which is {\bf non-ideal}.\\

One can utilize the classification into {\bf M}$_{\mathbf{x,y}}\$ complete \m s also for pre\m s if they are homogeneous in the sense that {\bf all} the final complete-\M components of the pre\M belong to one and the same kind {\bf M}$_{\mathbf{x,y}
}\$ of the 5 complete \m s. Those that lack such homogeneity can be classified by the worst final complete-\M component, "worst" meaning that it is farthest from the beginning in reading order on the array.\\

\noindent {\bf IX. SUMMING UP THE EQUIVALENT DEFINITIONS}\\

Let us {\bf sum up} that we have obtained
{\it 7 equivalent} definitions of {\bf general       pre\m } in {\bf section 2}
:

{\it 1)} the 'statistical form' of the \CC (4),

{\it 2)} its 'invariance form' (6),

{\it 3)} its 'strong invariance' form (11);

{\it 4)} the '\prc ' (10),

{\it 5)} the 'basic dynamical' characterization (8),

{\it 6)} the 'basis-dynamical' characterization (9a), and, as the other side of the coin, the 'subspace-dynamical' criterion (9b), and finally

{\it 7)} the 'canonical subsystem-basis expansion' criterion (13a) with (13b).

To the author's knowledge, new are 3), 5) and 7).\\

Let us sum up the results of {\bf section (4)}. We have defined {\bf nondemolition pre\M } {\it by additional requirements in 10 equivalent ways}. We have extended the 7 equivalent definitions of general       pre\m , and thus we have obtained:

{\it 1)} the 'extended statistical \cc ' definition (4) + (16a),

{\it 2)} the 'extended invariance \cc ' criterion (6) + (16b),

{\it 3)} the 'extended strong invariance' characterization (11) + (17),

{\it 4)} the 'extended \prc ' (10) + (24),

{\it 5)} the 'extended dynamical' definition (8) + (18),

{\it 6)} the 'extended basis-dynamical' characterization (9)+(19a) or equivalently (19b) by itself; further, the 'extended subspace-dynamical' condition (19c);

{\it 7)} the 'canonical subsystem-basis  expansion' criterion (25a) with (25b), and

{\it 8)} the 'twin-corre;ated canonical Schmidt decomposition' definition (26).

The 'canonical subsystem-basis expansion' characterization of general       pre\M (13a) with (13b) was extended in two ways (items 7) and 8)).

Two additional conditions without a counterpart in general       pre\M were given:

{\it 9)} The Pauli 'definition of repeatability' (23a) or (23b), and

{\it 10)} the 'twin-observables relation' (20).

As far as the author can tell, new are 3) and 8).

Characterization in item 10) had the consequences of lack of coherence (21a) in the final object state with respect to the measured observable \$O_A\$ and analogously (21b) regarding the final measuring-instrument state and the pointer observable \$P_B\$.\\

Finally, let us  sum up the three equivalent definitions of ideal pre\m :

{\it 1)} The canonical-final-state definition (32),

{\it 2)} the L\"uders-change-of-state definition (33a-c), and

{\it 3)} the strongly extended \CC definition (34a-c).\\

\noindent {\bf X. CONCLUDING REMARKS}\\

Let us conclude this review by a few remarks on some aspects that have been {\bf omitted}. What has been covered is outlined at the very beginning of the article.\\

{\bf (I)} Complete \M was viewed only as a constituent of pre\m .\\

{\it {\Large Literature} A quote from the first passage of the last chapter of the book on \QM by Peres (2002) reads:

"In order to observe a physical system, we make it interact with an apparatus. The latter must be described by quantum mechanics, because there is no consistent dynamical
scheme in which a quantum system interacts with a classical one. On the other
hand, the result of the observation is
recorded by the apparatus in a classical
form ..."

Transition from \QMl description of the \MI to that of classical physics Peres calls "dequantization". It seems to be way out of the scope of unitary \Q \M theory. See also Hay, and Peres, 1998.}\\

ng other things, the expounded theory {\bf allows generalization} (extension) in a number of its basic concepts that have not been covered.\\

{\bf (II)} Both the initial state of the object \$\ket{\phi}_A\IN\$ and that of the measuring instrument \$\ket{\phi}_B\IN \$ allow generalization to {\bf general states} (density operators) \$\rho_A\IN\$ and \$\rho_B\IN \$ respectively.

It is to be expected that, as it is usually the case with density operators, the relations valid for general states can easily be obtained from the corresponding relations valid for pure states when one decomposes the general states into pure ones. Besides, there is so-called purification: Every density operator can be viewed as the reduced density operator of a bipartite pure state.\\

{\it {\Large Literature} As to general states, von Neumann has proved in his famous no-go theorem (von Neumann, 1955, first part of section 3. in chapter VI) that unitary \M theory cannot explain complete \m , i. e., the fact that the individual systems end up in one branch $$F_B^k\ket{\Phi}_{AB}\f\Big/
||F_B^k\ket{\Phi}_{AB}\f||\eqno{(46)}$$ of the final pre\M state \$\ket{\Phi}_{AB}\f\$ (if the complete \M is a minimal one, cf section V). We outline the claim of this theorem.

In a purely pure-state \M theory, as in the present review, it is clear that the coherence \$\ket{\phi}_A\IN = \sum_kE_A^k\ket{\phi}_A\IN\$ with respect to the eigenvalues \$o_k\$ of the measured observable \$O_A\enskip\Big(=\sum_ko_kE_A^k\Big)\$ in the initial state \$\ket{\phi}_A\IN\$ of the object is not destroyed; it is only transformed into coherence $$\ket{\Phi}_{AB}^f= \sum_kF_B^k\ket{\Phi}_{AB}^f\eqno{(47a)}$$ with respect to the pointer observable \$P_B\enskip\Big(=\sum_kp_kF_B^k\Big)\$  in the final pre\M state \$\ket{\Phi}_{AB}^f\$ (cf (15) for the separate evolution of each branch). Though von Neumann did not expound a detailed theory of pre\m , he did not
consider the \M paradox in the pure state case because the unitary evolution operator takes the pure composite state
\$\ket{\phi}_A\IN\ket{\phi}_B\IN\$ into a pure state \$\ket{\Phi}_{AB}^f\$, and there is no way how coherence could disappear.

What von Neumann did was to assume that the initial state of the measuring instrument is in a mixed state \$\rho_B\IN\$; a proper mixture due to incomplete knowledge about the state. Then one might {\it conjecture} that this mixture would lead to the mixture
$$\rho_{AB}\f =\sum_k\bra{\phi}_A\IN E_A^k\ket{\phi}_A\IN\times$$ $$
\Big(F_B^k\ket{\Phi}_{AB}\f\Big/
||F_B^k\ket{\Phi}_{AB}\f||\Big)\Big(
\bra{\Phi}_{AB}\f F_B^k\Big/
||F_B^k\ket{\Phi}_{AB}\f||\Big)\eqno{(47b)}$$ (cf (14)), which is observed  in the laboratory. Von Neumann disproves in detail this conjecture.

Von Neumann's no-go theorem has been often considered as a proof of the \M paradox,
i.e., of the puzzle why unitary \QM does not furnish separate branches (terms in (47b)) for the individual measured objects. In particular, it is puzzling how a deterministic theory, as the one reviewed in this article, can lead to the indeterministic branches for the individual objects, which becomes deterministic (described by (47b)) on the ensemble level.

No other conceivable way (than the one treated in von Neumann's no-go theorem) how one could obtain (47b) within unitary dynamics of \M was seen till Everett 1957 and 1973 and De Witt 1973 shocked the world by hypothesizing that the separate branches of pre\M might become parallel worlds and that we, and everything that we know, somehow become one of these worlds.}\\

{\bf (III)} The discrete observables \$O_A=\sum_ko_kE_A^k\$ to which this review has been confined can be generalized to include also observables that contain a continuous part in their spectrum, in particular purely continuous observables as position and linear momentum.\\

{\it {\Large Literature} Busch and Lahti (1987), and also Busch,  Grabowski, and Lahti (1995a) have investigated this subject.}\\

{\bf (IV)} The concept of observables that are measured can be extended from ordinary ones to ones that are expressed as POV (positive-operator valued) measures (cf Busch, Lahti, and Mittelstaedt, 1996). The PV (projector-valued) measures corresponding to ordinary observables are special cases.\\

{\it {\Large Literature} One should read
section 3.6 in Part I of Vedral (2006).
Generalized observables (POV measures) are described in detail in the book by Busch, Grabowski, and Lahti (1995b). Also the study in Busch, Kiukas, and Lahti (2008, section 3.), on connection between POV and PV measures based on the Neumark (1940) dilation theorem is recommended. Also the article Peres (1990) is relevant and interesting.}\\

{\bf (V)} The eigen-projectors \$E_A^k\$ of an ordinary observable can be generalized by positive operators, the physical meaning of which is 'effects'.
The projectors, with the meaning of events or properties or \Q statements, are special cases.

In the most general case, which is studied in Busch, Lahti, and Mittelstaedt 1996, pre\M is defined by the \prc , and generally this is not equivalent to the \cc . The present investigation was undertaken in the hope that restriction to the physically most important case of ordinary discrete observables will help to delve deeper into the subject, obtain more results, and see them with more clarity and simplicity.\\

{\bf (VI)} The theory presented in the present review is purely algebraic. The question of {\bf feasibility} of the particular pre\M procedures is not discussed.\\

{\it {\Large Literature} Concerning this important aspect of \m , one should read Cassinelli and Lahti (1990) and section 6. in chapter III of Busch, Lahti, and Mittelstaedt 1996. One may also learn about the Wigner-Yanase-Araki theorem, which claims to set serious limitations on what can be measured exactly. One may read the critical short article Ohira and Pearle (1988), and the references therein.}\\

{\bf (VII)} Measurements cannot be performed without {\bf preparation}, a procedure that brings about the initial states \$\ket{\phi}_A\IN\$.\\

{\it {\Large Literature} One can read about preparation in section 8. of chapter III in Busch, Lahti, and Mittelstaedt 1996 or in Herbut 2001.}\\

{\bf (VIII)} {\bf Information-theoretical aspects} of \M theory have not been discussed in the present review either.\\

{\it {\Large Literature} Information gain in various kinds of \M is investigated in Lahti,  Busch, and Mittelstaedt  1991. One can read about this aspect in section 4. of chapter III of Busch, Lahti, and Mittelstaedt 1996. Entropic, information-theoretical and coherence aspects of \M were studied in a number of articles by the present author Herbut (2002, 2003a, 2003b, and 2005).}\\

\vspace{1cm}
I am grateful to my onetime associates: the late Milan Vuji\v{c}i\'c, further
Milan Damnjanovi\'c, Igor Ivanovi\'c, and Maja Buri\'c for helpful and inspiring discussions on \M theory.\\

\noindent {\bf Appendix Proof of an auxiliary algebraic certainty claim}\\

We {\it prove} now the general claim that the following equivalence is valid for a pure state \$\ket{\psi}\$ and an event \$E\$:
$$\bra{\psi}E\ket{\psi}=1\quad\Leftrightarrow\quad
\ket{\psi}=E\ket{\psi}.$$ Evidently
$$\bra{\psi}E\ket{\psi}=1
\quad\Rightarrow\bra{\psi}E^c\ket{\psi}=0,$$
where \$E^c\equiv I-E\$ is the ortho-complementary projector and \$I\$ is the identity operator. Further, one has \$||E^c\ket{\psi}||=0\$,  \$E^c\ket{\psi}=0\$, and \$E\ket{\psi}=\ket{\psi}\$ as claimed. The inverse implication is obvious.\\

\noindent
{\bf REFERENCES}\\

\noindent
Beltrametti, E. G., G. Cassinelli, and P. J. Lahti, 1990,

\indent "Unitary Premeasurements of Discrete Quantities in

\indent Quantum Mechanics", J. Math. Phys., {\bf 31}, 91.

\noindent
Bohm, D., 1952, {\it Quantum Theory}, (Prentice-Hall Inc.,New

\indent
York).

\noindent
Busch, P., and P. J. Lahti, 1987,
"Minimal Uncertainty

\indent and Maximal Information for Quantum Position and

\indent Momentum", J. Phys. A: Math. Gen. {\bf 20}, 899.

\noindent
Busch, P., M. Grabowski, and P. J. Lahti, 1995a, "Re-

\indent peatable Measurements in Quantum Theory: Their

\indent Role and Feasibility," Found. Phys., {\bf 25}, 1239.

\noindent
Busch, P., M.  Grabowski, and P. J. Lahti, 1995b, {\it Ope-

\indent rational Quantum Physics} (Springer, Berlin).

\noindent
Busch, P., and P. J. Lahti, 1996,
"Correlation Properties

\indent of Quantum Measurements", J.Math.Phys., {\bf 37},  2585.

\noindent
Busch, P., P. J. Lahti, and P. Mittelstaedt, 1996, {\it The

\indent Quantum Theory of Measurement}, 2nd edition

\indent (Springer, Berlin).

\noindent
Busch, P., J. Kiukas, and P. J. Lahti, 2008, "Measuring

\indent Position and Moment Together" Phys. Lett. A, {\bf 372},

\indent 4379.

\noindent
Cassinelli, G., and P. J. Lahti, 1990, "Strong-Correlation

\indent Measurements in Qunatum Mechanics", Il Nuovo Ci-

\indent mento, {\bf 105}, 1223.

\noindent
Cohen-Tannoudji, C., B. Diu, and F. Laloe, 1977, {\it Quan-

\indent tum Mechanics} vol. I (Wiley-Interscience, New York),

\indent p. 221 (Fifth Postulate).

\noindent
D'Espagnat, B., 1976, {\it Conceptual Foundations of Quan-

\indent tum Mechanics}, Second Edition (W. A. Benjamin, Inc.,

\indent Reading,
Massachusetts) subsection 7.2 .

\noindent
De Witt B. S., 1973, "The Many-Universes Interpretation

\indent of Quantum Mechanics," in: {\it The Many-Worlds Inter-

\indent pretation of
Quantum Mechanics}, edited by B. De Witt

\indent and N. Graham (Princeton University Press, Prince-

\indent ton), pp. 167-219.

\noindent
Einstein, A., Podolsky, B., and Rosen, N., 1935,
"Can

\indent
Quantum-Mechanical Description of
Reality Be Considered

\indent
Complete?", Phys. Rev., {\bf 47}, 777.

\noindent
Everett, H., 1957, ""Relative State" Formulation of

\indent Quantum Mechanics," Rev. Mod. Phys., {\bf 29}, 454.

\indent Reprinted in {\it The Many-Worlds Interpretation of

\indent Quantum Mechanics}, edited by B. De Witt and N. Gra-

\indent ham, 1973 (Princeton University Press, Princeton), pp.

\indent 141-150.

\noindent
Everett, H., 1973, "The Theory of the Universal Wave

\indent Function," in: {\it The Many-Worlds Interpretation of

\indent Quantum Mechanics}, edited by B. De Witt and N. Gra-

\indent ham (Princeton University Press, Princeton), pp. 1-

\indent 140.

\noindent
Hay, O., and A. Peres, 1998,
"Quantum and Classical

\indent Descriptions of a Measuring Apparatus," Phys. Rev.

\indent A, {\bf 58}, 116.

\noindent
Herbut, F., 1969,  "Derivation
of the Change of State

\indent in Measurement
from the Concept of Minimal Mea-

\indent surement," Ann. Phys. (N. Y.), {\bf 55}, 271.

\noindent
Herbut, F., 1974, "Minimal-Disturbance Measurement

\indent as a Specification in von Neumann's Quantal Theory

\indent of Measurement," Intern. J. Theor. Phys., {\bf 11},  193.

\noindent
Herbut, F., 2001, "A Theory of Quantum Prepara-

\indent tion and the
Corresponding Advantage of the Relative-

\indent Collapse
Interpretation of Quantum Mechanics as

\indent Compared to the Conventional One," arxiv:quant-

\indent ph/0107064.

\noindent
Herbut, F., 2002, "Chains of Quasiclassical Information

\indent for Bipartite Correlations and the Role of Twin Ob-

\indent servables," Phys. Rev. A, {\bf 66}, 052321; also December

\indent issue Virt. J. Quant.  Inf.

\noindent
Herbut, F., 2003a, "The Role of Coherence Entropy of

\indent Physical Twin Observables in Entanglement," J. Phys.

\indent A: Math. Gen., {\bf 36}, 8479.

\noindent
Herbut, F., 2003b, "On the Meaning of Entanglement in

\indent Quantum Measurement," arxiv:quant-ph/0311192.

\noindent
Herbut, F., 2004, "Distinguishing Quantum Measure-

\indent ments of Observables
in Terms of State Transformers,"

\indent arxiv:quant-ph/0403101.

\noindent
Herbut, F., 2005, "A Quantum Measure of Coherence

\indent and Incompatibility" J. Phys. A: Math. Gen., {\bf 38},

\indent 2959.

\noindent
Herbut, F., 2007a, "Quantum Probability Law from

\indent 'Environment-Assisted
Invariance' in Terms of

\indent Pure-State Twin Unitaries," J. Phys. A: Math. Theor.,

\indent  {\bf 40}, 5949.

\noindent
Herbut, F., 2007b, "Derivation of the Quantum Probabi-

\indent lity Law from
Minimal Non-Demolition Measure-

\indent ment," J. Phys. A: Math. Theor., {\bf 40},  10549.

\noindent
Herbut, F., 2014a, "Fleeting Critical Review of the

Recent Ontic Breakthrough in Quantum Mechanics,"

\indent arxiv:1409.6290 [quantum-ph].

\noindent
Herbut, F., 2014b, "Bipartite Entanglement
Review of

\indent
Subsystem-Basis Expansions
and Correlation Operators

\indent
in It", arxiv:1410.4988 [quant-ph].

\noindent
Herbut, F., 2014c, "Subsystem Measurement in Unitary
\indent Quantum Measurement Theory with Redundant Entan-

\indent glement," Int. J. Quant. Inf., {\bf 12},  1450032 (16 pages);

\indent arxiv:1302.2250 [quant. phys.].

\noindent
Herbut, F., and M. Vuji\v{c}i\'{c}, 1976, "Distant Measure-

\indent ment", Ann. Phys. (New York), {\bf 96}, 382.

\noindent
Khrennikov, A., 2008, "The Role of Von Neumann and

\indent L\"uders Postulates in the Einstein, Podolsky, and Rosen

\indent Considerations: Comparing Measurements with De-

\indent generate and Nondegenerate Spectra,"  J. Math. Phys. {\bf

\indent 49},  052102.

\noindent
Khrennikov, A., 2009a, "Bell's Inequality, Einstein,

\indent Podolsky, Rosen
Arguments and Von Neumann's Pro-

\indent jection Postulate," Laser Phys., {\bf 19}, 346.

\noindent
Khrennikov, A., 2009b, "Von Neumann and L\"uders Pos-

\indent tulates and Quantum Information Theory," Int. J.

\indent Quant. Inf., {\bf 7}, 1303.

\noindent
Kraus, K., 1983, {\it States, Effects, and Operations}

\indent (Springer-Verlag, Berlin).

\noindent
Lahti, P. J., 1990, "Quantum Theory of Measurement

\indent and the Polar Decomposition of an Interaction", Int.

\indent J. Theor. Phys., {\bf 29}, 339.

\noindent
Lahti, P. J., 1993, "Quantum Measurement II," Int. J.

\indent Theor. Phys., {\bf 32}, 1777, International Quantum Struc-

\indent tures Association, biannual meeting 1992, Castinglion-

\indent cello, Italy, eds. E. Beltrametti, G. Cattaneo, M.L.

\indent Dalla Chiara.

\noindent
Lahti, P. J., P. Busch, and P. Mittelstaedt,  1991, "Some

\indent Important Classes of Quantum Measurements and

\indent Their Information Gain," {\bf 32}, 2770.

\noindent
Leifer, M. S., 2014, "Is the Quantum State Real? A

\indent Review of Psi-Ontology Theorems', arxiv:1409.1570v1

\indent [quant-ph].

\noindent
L\"{u}ders, G., 1951, "\"Uber die Zustands\"anderung durch den

\indent Messprozess,"  Ann. der Physik,  {\bf 8}, 322.

\noindent
Messiah, A., 1961, {\it Quantum Mechanics} vol. I, (North

\indent Holland, Amsterdam), p 333.

\noindent
Neumark, M. A., 1940, Izv. Akad. Nauk SSSR, Ser.

\indent Mat. {\bf 4}, 53, 277; 1943, C.R. (Doklady) Acad. Sci.

\indent URSS (N.S.) {\bf 41}, 359.

\noindent
Ohira, T., and Ph. Pearle, 1988, "Perfect Disturbing

\indent Measurements", {\bf 56}, 692.

\noindent
Ozawa, M., 1984, "Quantum Measuring Processes of

\indent Continuous Observables," J. Math. Phys., {\bf 25}, 79.

\noindent
Pauli, W., 1933, {\it Die Allgemeinen Prinzipien der Wellen-

\indent mechanik},  Handbuch der Physik, edited by H. Geiger

\indent and K. Scheel, 2nd edition, Vol. 24, (Springer-Verlag,

\indent Berlin) pp. 83-272. English translation: {\it General Prin-

\indent ciples of Quantum Mechanics}, 1980 (Springer-Verlag,

\indent Berlin).

\noindent
Peres, A., 1974, "Quantum Measurements are Re-

\indent versible," Am. J. Phys., {\bf 42}, 886.

\noindent
Peres, A., 1990, "Neumark's Theorem and Quantum In-

\indent separability", Found. Phys., {\bf 20}, 1441.

\noindent
Peres, A., 2002,  {\it Quantum Theory: Concepts and Methods}

\indent (Kluwer Academic Publishers, New York).

\noindent
Pusey, M. F., J. Barrett, and T. Rudolph, 2012, "On the

\indent Reality of the Quantum State", Nature Phys. {\bf 8}, 475;

\indent arXiv:1111.3328 .

\noindent
Schr\"{o}dinger, E., 1935, "Discussion of Probability Rela-

\indent tions Between Separated Systems," Proc. Cambridge

\indent Phil. Soc., {\bf 31}, 555.

\noindent
Vedral, V., 2006, {\it Introduction to Quantum Information

\indent Science} (Oxford University Press, Oxford, United

\indent Kingdom).

\noindent
Von Neumann, J., 1955,{\it  Mathematical Foundations

\indent of Quantum Mechanics},
(Princeton University Press,

\indent Princeton).

\end{document}